\RequirePackage{fix-cm}

\documentclass[twocolumn]{svjour3}          
\smartqed  
\usepackage{xcolor}
\usepackage{graphicx}
\usepackage{amssymb}
\usepackage{amsmath}
\usepackage{subcaption}
\usepackage{multirow}
\usepackage{hyperref}

\usepackage[table ]{ xcolor}
\begin{document}

\title{Towards Robust and Realible Multimodal Misinformation Recognition with Incomplete Modality
}


\author{Hengyang Zhou         \and
        Yiwei Wei \and Jian Yang \and Zhenyu Zhang
}


\institute{Hengyang Zhou \at
              Nanjing University \\
              \email{hengyangzhou@smail.nju.edu.cn}           
           \and
           Yiwei Wei \at
              China University of Petroleum \\
              \email{weiyiwei@cupk.edu.cn}
            \and 
                Jian Yang \at
                Nanjing University of Science and Technology\\
                \email{csjyang@njust.edu.cn}
            \and
            Zhenyu Zhang \at
            Nanjing University \\
            \email{zhenyuzhang@nju.edu.cn}\\
            Corresponding author
}

\date{Received: date / Accepted: date}

\maketitle

\begin{abstract}
Multimodal Misinformation Recognition has become an urgent task with the emergence of huge multimodal fake content on social media platforms. Previous studies mainly focus on complex feature extraction and fusion to learn discriminative information from multimodal content. However, in real-world applications, multimedia news may naturally lose some information during dissemination, resulting in modality incompleteness, which is detrimental to the generalization and robustness of existing models. To this end, we propose a novel generic and robust multimodal fusion strategy, termed Multi-expert Modality-incomplete Learning Network (MMLNet), which is simple yet effective. It consists of three key steps: (1) Multi-Expert Collaborative Reasoning to compensate for missing modalities by dynamically leveraging complementary information through multiple experts. (2) Incomplete Modality Adapters compensates for the missing information by leveraging the new feature distribution. (3) Modality Missing Learning leveraging an label-aware adaptive weighting strategy to learn a robust representation with contrastive learning. We evaluate MMLNet on three real-world benchmarks across two languages, demonstrating superior performance compared to state-of-the-art methods while maintaining relative simplicity. By ensuring the accuracy of misinformation recognition in incomplete modality scenarios caused by information propagation, MMLNet effectively curbs the spread of malicious misinformation. Code is publicly available at \textcolor{blue}{https://github.com/zhyhome/MMLNet}.

\keywords{Incomplete Modality \and Misinformation Recognition \and Robustness Enhancement \and Multimodal Learning}
\end{abstract}

%


\section{Introduction}
With the widespread use of social media platforms like Twitter and Weibo, users can easily share personal opinions, emotions, and daily information. However, the misuse of these platforms and the lack of effective regulation have led to the rapid spread of misinformation, making the issue of misinformation  increasingly severe \cite{survey_rumours}. This phenomenon has profound social impacts and poses significant threats to areas such as politics, finance, and public health. The abuse of advanced generative models has further exacerbated the problem, contributing to the rise of text-based fake news \cite{text_fake} and visual deepfakes \cite{visual_fake}. In addition, the dissemination of multimodal forged media has increased the reach of false information, increasing its potential to mislead the public \cite{fkaowl}. Given the coexistence of fake information in both the image and text modalities, detecting such misinformation presents an unprecedented challenge.

To address this challenge, current Multimodal Misinformation Recognition (MMR) methods \cite{fkaowl}, \cite{MOE_WWW25}, \cite{MMDFND}, \cite{2025-efficient-COLING} primarily focus on the fusion of cross-modal features. Despite the significant achievements of these methods, they do not consider scenarios involving incomplete modalities due to missing modality information in real-world settings, which introduces risks for practical applications.

\begin{figure}[h]
  \centering
  \includegraphics[width=\linewidth]{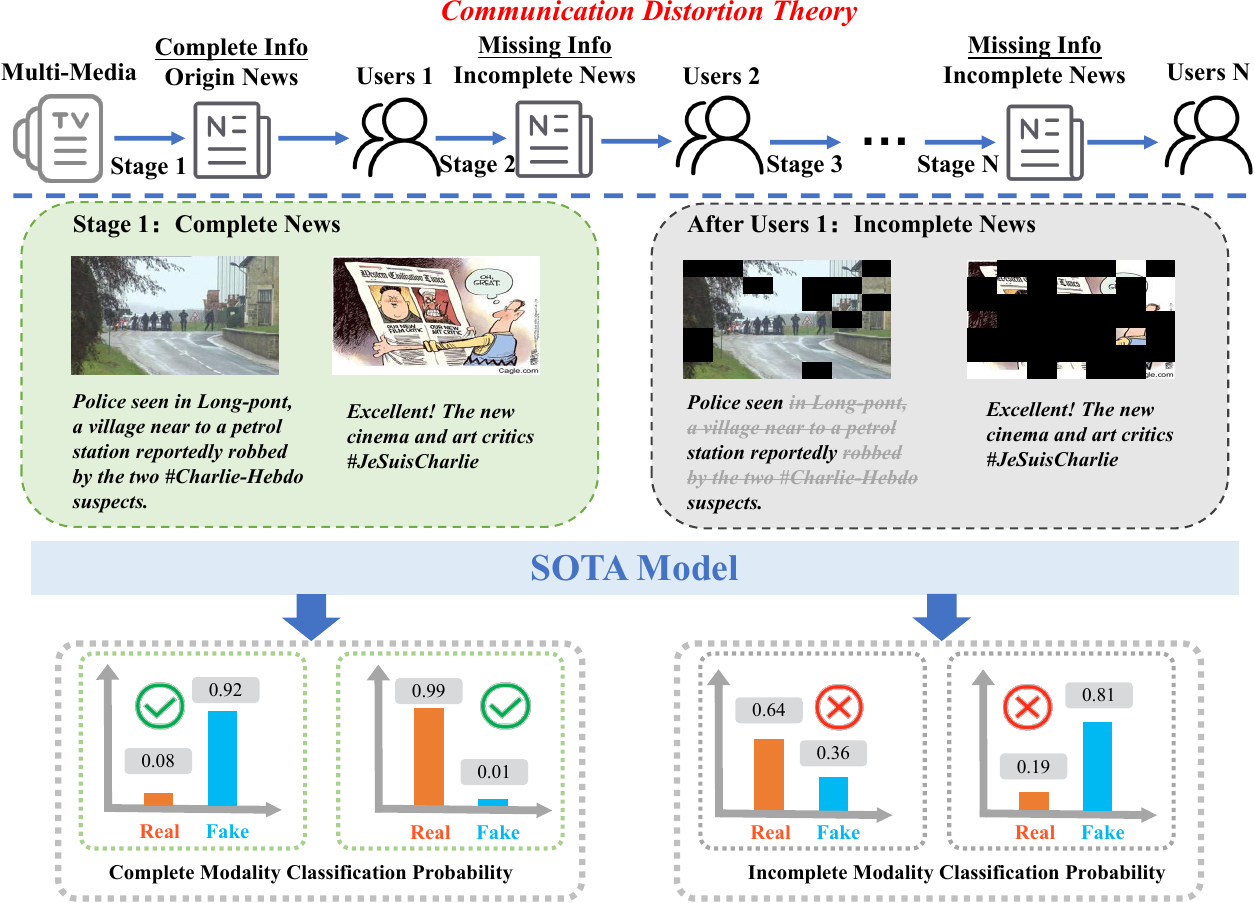}
  \caption{Illustration of misinformation scenarios with incomplete modalities. Existing works have only investigated the scenario of complete modalities in the first step of the distortion propagation theory.}
  \label{fig:fig1}
\end{figure}

According to the Communication Distortion Theory, during the process of information transmission, the information may undergo distortion, misunderstanding, or loss, resulting in incomplete or inaccurate information. Early studies \cite{data_missing_theory} have demonstrated this theory through behavioral experiments. \cite{DGAmm24}, \cite{10750410} also demonstrates that early recognition in the propagation path helps identify misinformation.

Figure \ref{fig:fig1} visually illustrates the application of this theory in the context of news dissemination. In Stage 1, multi-media initially publishes the original news. However, after passing through Group Users 1, the information may become distorted. For simplicity, we consider only the case of information loss. As a result, Group Users 2 will receive incomplete news, and different user groups will obtain news with varying degrees of completeness. Unfortunately, existing MMR methods are limited to verifying the authenticity of news at the first stage, relying on complete news data in both training and testing while overlooking the fact that multiple user groups receive incomplete multimodal news information. This limitation poses risks in real-world applications. For the first time, we address this issue and propose the Multi-expert Modality-incomplete Learning Network (MMLNet), which mitigates this phenomenon by incorporating incomplete multimodal data during both training and testing.

\definecolor{mydarkgreen}{RGB}{0, 128, 0}
\begin{table}
\setlength{\tabcolsep}{1.5pt}
  \caption{Part of experimental results show that the Accuracy score (\%) of existing SOTA models has declined in the face of incomplete modalities.  Incomplete modalities are dominated by incomplete text modalities and incomplete image modalities in this case. Note that `-' indicates that the modality is complete, and `↓' indicates the percentage of modality missing. Green fonts show the performance degradation of the model compared to the complete modality scenario.}
  \label{tab:DROP_xinneng}
\begin{tabular}{ccllllll}
\hline
\multicolumn{2}{c}{Modality} & \multicolumn{2}{c}{} & \multicolumn{4}{c}{Dataset} \\ \cline{1-2} \cline{5-8} 
Text & Image & \multicolumn{2}{c}{\multirow{-2}{*}{Method}} & \multicolumn{2}{l}{Weibo} & \multicolumn{2}{l}{Pheme} \\ \hline
 &  & \multicolumn{2}{l}{NSLM(AAAI'24)} & \multicolumn{2}{l}{92.28} & \multicolumn{2}{l}{84.65} \\
 &  & \multicolumn{2}{l}{MIMoE(WWW'25)} & \multicolumn{2}{l}{92.49} & \multicolumn{2}{l}{85.64} \\
 &  & \multicolumn{2}{l}{Qwen-VL-7B} & \multicolumn{2}{l}{81.71} & \multicolumn{2}{l}{55.12} \\
 &  & \multicolumn{2}{l}{LLaVA-13B} & \multicolumn{2}{l}{83.00} & \multicolumn{2}{l}{57.10} \\
\multirow{-5}{*}{-} & \multirow{-5}{*}{-} & \multicolumn{2}{l}{\cellcolor[HTML]{ECF4FF}MMLNet(ours)} & \multicolumn{2}{l}{\cellcolor[HTML]{ECF4FF}95.22} & \multicolumn{2}{l}{\cellcolor[HTML]{ECF4FF}87.78} \\ \hline
 &  & \multicolumn{2}{l}{NSLM(AAAI'24)} & \multicolumn{2}{l}{86.07 \textcolor{mydarkgreen}{\scriptsize↓6.21}} & \multicolumn{2}{l}{82.50 \textcolor{mydarkgreen}{\scriptsize↓2.15}} \\
 &  & \multicolumn{2}{l}{MIMoE(WWW'25)} & \multicolumn{2}{l}{90.85 \textcolor{mydarkgreen}{\scriptsize↓1.64}} & \multicolumn{2}{l}{77.88 \textcolor{mydarkgreen}{\scriptsize↓7.76}} \\
 &  & \multicolumn{2}{l}{Qwen-VL-7B} & \multicolumn{2}{l}{72.90 \textcolor{mydarkgreen}{\scriptsize↓8.81}} & \multicolumn{2}{l}{45.87\footnotesize \textcolor{mydarkgreen}{\scriptsize↓9.25}} \\
 &  & \multicolumn{2}{l}{LLaVA-13B} & \multicolumn{2}{l}{76.72 \textcolor{mydarkgreen}{\scriptsize↓6.28}} & \multicolumn{2}{l}{48.35 \textcolor{mydarkgreen}{\scriptsize↓8.75}} \\
\multirow{-5}{*}{↓25(\%)} & \multirow{-5}{*}{↓75(\%)} & \multicolumn{2}{l}{\cellcolor[HTML]{ECF4FF}MMLNet(ours)} & \multicolumn{2}{l}{\cellcolor[HTML]{ECF4FF}90.23 \textcolor{mydarkgreen}{\scriptsize↓4.99}} & \multicolumn{2}{l}{\cellcolor[HTML]{ECF4FF}82.83 \textcolor{mydarkgreen}{\scriptsize↓4.95}} \\ \hline
\multicolumn{1}{l}{} & \multicolumn{1}{l}{} & \multicolumn{2}{l}{NSLM(AAAI'24)} & \multicolumn{2}{l}{71.74 \textcolor{mydarkgreen}{\scriptsize↓20.54}} & \multicolumn{2}{l}{74.09 \textcolor{mydarkgreen}{\scriptsize↓10.56}} \\
\multicolumn{1}{l}{} & \multicolumn{1}{l}{} & \multicolumn{2}{l}{MIMoE(WWW'25)} & \multicolumn{2}{l}{80.06 \textcolor{mydarkgreen}{\scriptsize↓12.43}} & \multicolumn{2}{l}{74.25 \textcolor{mydarkgreen}{\scriptsize↓11.39}} \\
\multicolumn{1}{l}{} & \multicolumn{1}{l}{} & \multicolumn{2}{l}{Qwen-VL-7B} & \multicolumn{2}{l}{65.46 \textcolor{mydarkgreen}{\scriptsize↓16.25}} & \multicolumn{2}{l}{42.08 \textcolor{mydarkgreen}{\scriptsize↓13.04}} \\
\multicolumn{1}{l}{} & \multicolumn{1}{l}{} & \multicolumn{2}{l}{LLaVA-13B} & \multicolumn{2}{l}{67.99 \textcolor{mydarkgreen}{\scriptsize↓15.01}} & \multicolumn{2}{l}{43.56 \textcolor{mydarkgreen}{\scriptsize↓13.54}} \\
\multicolumn{1}{l}{\multirow{-5}{*}{↓75(\%)}} & \multicolumn{1}{l}{\multirow{-5}{*}{↓25(\%)}} & \multicolumn{2}{l}{\cellcolor[HTML]{ECF4FF}MMLNet(ours)} & \multicolumn{2}{l}{\cellcolor[HTML]{ECF4FF}92.55 \textcolor{mydarkgreen}{\scriptsize↓2.66}} & \multicolumn{2}{l}{\cellcolor[HTML]{ECF4FF}80.19 \textcolor{mydarkgreen}{\scriptsize↓7.59}} \\ \hline
\end{tabular}
\end{table}

Although existing MMR approaches have achieved significant success, their performance degrades substantially in highly incomplete modality scenarios. This decline occurs because these methods overly emphasize multimodal fusion representations while neglecting the joint representation of unimodal information. Consequently, when unimodal information is missing, it further disrupts the effectiveness of multimodal fusion representations. We conduct comparative experiments on complete and incomplete modalities using the latest publicly released state-of-the-art open-sourced MMR methods and finetuned advanced Multimodal Large Language Models (MLLM), including NSLM \cite{NSLMaaai24}, MIMoE \cite{MOE_WWW25}, Qwen-VL \cite{Qwen-VL}, and LLaVA \cite{liu2023llava}, on real-world datasets in two different languages, as shown in Table \ref{tab:DROP_xinneng}.

In multimodal learning, extensive research has been conducted on the handling of incomplete modalities \cite{MMAlign_emnlp22}, \cite{Sun2022EfficientMT}, \cite{MM24robustSenDistribution}, \cite{MM24robustSenMOE}, \cite{robustSenMM21}.  Existing approaches in incomplete multimodal learning \cite{MM24robustSenDistribution}, \cite{MM24robustSenMOE}, \cite{robustSenMM21} primarily focus on reconstructing missing modalities from available ones \cite{nips24robustSen1}, \cite{recover4}, aiming to maintain feature consistency across samples with arbitrary modality gaps. While these techniques have shown promising results, their direct application to image-text pair based MMR presents two fundamental challenges. First, existing methods are typically developed independently for different problems, leading to heterogeneous models with distinct design strategies. This architectural divergence complicates direct integration and increases the alignment burden. Second, misinformation in social media often results in semantic inconsistencies between images and text, which contradicts the widely held assumption in domains such as video and medical imaging \cite{GCNetTPAMI}, \cite{translationAAAI19}, \cite{nips23robustEmo} that modalities originating from the same sample are inherently correlated. This lack of cross-modal coherence hinders the effectiveness of conventional approaches in MMR, necessitating novel solutions tailored to its unique challenges.

The most related work is \cite{IMOL}, however, which focuses on detecting the authenticity of video modalities and is coarse-grained for the absence of incomplete modalities (missing the entire modality), which is not suitable for the scenario studied in this paper.

To fill this gap, we first investigate the impact of incomplete modalities on MMR and propose MMLNet. The main contributions of this paper can be summarized as follows:
\begin{itemize}
\item We focus on robust multimodal misinformation  recognition under incomplete modalities, an aspect that has been less emphasized in previous research but is prevalent in real-world scenarios. To the best of our knowledge, this is a pioneering exploration of model robustness in MMR.
\item We propose MMLNet, a novel approach to handle arbitrary incomplete modality information for MMR tasks in a unified framework. It leverages multi-expert learning of feature distributions of various modalities, compensates for missing modality information through incomplete modality adapters, and further employs modality missing learning to enhance modality distribution for robust MMR.
\item Experimental results show that MMLNet achieves significant improvements over sota MMR methods and MLLM on three real-world MMR datasets across two languages, validating its superiority in robust MMR.
\end{itemize}

\section{Related Work}
\subsection{Misinformation recognition}
\subsubsection{Unimodal Methods}
Text-based approaches focus on different analytical dimensions. Some methods integrate social context by leveraging propaganda strategies \cite{huang-etal-2023-faking}, user feedback \cite{10.1145/3485447.3512163}, temporal trends \cite{hu-etal-2023-learn}, and news environments \cite{sheng-etal-2022-zoom}. Ghanem et al. \cite{ghanem-etal-2021-fakeflow} proposed a technique that combines topic and sentiment features extracted from textual data. More recently, Nan et al. \cite{weibo21} and Zhu et al. \cite{10.1109/TKDE.2022.3185151} explored domain shift issues caused by variations in word frequency and sentiment, introducing domain gating and domain memory mechanisms to improve domain representation.

Various image-based methods \cite{Yang2022ConfidenceCalibratedFI} have been introduced to assess the authenticity of visual content by analyzing edited traces. Some CNN-based techniques operate in the spatial domain to identify artifact patterns, utilizing approaches such as blending \cite{Li_2020_CVPR_fake}, multiple instance learning \cite{10.1145/3394171.3414034}, patch consistency analysis \cite{Zhao_2021_ICCV_fake}, reconstruction techniques \cite{10.1145/3394171.3413732}, \cite{10.1007/978-3-031-19781-9_8}, and local feature mining \cite{Dong2022ImplicitIL}. Alternatively, other studies focus on transforming images into the frequency domain through methods like discrete cosine transform \cite{Qian2020ThinkingIF}, composite phase spectrum analysis \cite{Liu2021SpatialPhaseSL}, and high-frequency noise extraction \cite{Luo2021GeneralizingFF}.

\subsubsection{Multimodal Methods}
Mainstream multimodal MMR methods extract semantic representations by integrating cross-modal textual and visual features \cite{10.1145/3123266.3123385}. Some approaches enhance semantic information by incorporating external knowledge bases, as proposed by Sabir et al. \cite{10.1145/3240508.3240707} and Wang et al. \cite{10.1145/3372278.3390713}. Qi et al. \cite{10.1145/3474085.3481548} focused on extracting visual entities to improve high-level news comprehension. Peng et al. \cite{weibo} applied named entity recognition and word segmentation to enhance Chinese social media identification. To better fuse textual and visual features, co-attention networks \cite{wu-etal-2021-multimodal} and contextual attention networks \cite{10.1145/3404835.3462871} have been developed. Kochkina et al. \cite{pheme} employed a multi-task learning approach for rumor verification. Additionally, Ying et al. \cite{10.1609/aaai.v37i4.25670} introduced an improved Mixture of Experts network to refine feature integration across multiple perspectives. To mitigate spurious correlations from modality inconsistencies and data biases, ambiguity learning \cite{10.1145/3485447.3511968} and causal inference \cite{chen-etal-2023-causal} were introduced separately. FAK-Owl \cite{fkaowl} performs knowledge-enhanced verification based on Large Vision-Language Models. NSLM \cite{NSLMaaai24} utilizes neuro-symbolic reasoning for misinformation recognition. \cite{yin2024fine} proposed a method based on heterogeneous graphs for graph attention networks, achieving fine-grained multimodal deepfake classification. MIAN \cite{zhang2025multimodalinverseattentionnetwork} is an attention network based on the utilization of intrinsic discriminative features. DGA-FAKE \cite{DGAmm24} detects misinformation early through the diffusion propagation path. Recent work MIMoE \cite{MOE_WWW25} detects misinformation through interactive mixture of experts. 

These methods effectively utilize image and text modalities to uncover implicit misinformation but struggle to handle missing modalities. Since such issues frequently occur in real-world applications \cite{data_missing_theory}, we propose MMLNet, a robust and practical approach for detecting misinformation in news.

\subsection{Robust Multimodal Learning}
The issue of missing modalities is prevalent across all types of multimodal data and has been widely studied. Early research employed data imputation methods \cite{GCNetTPAMI} to recover missing modalities from available ones. Some approaches directly fill in missing modalities with fixed values \cite{10.1145/3394486.3403182}, \cite{9258396}. Additionally, some researchers leverage the generative capabilities of specific neural network architectures, such as autoencoders \cite{10.1145/1390156.1390294} and Transformers \cite{10.5555/3295222.3295349}. \cite{guo-etal-2024-multimodal}, \cite{hu2024deep},  \cite{lee2023cvpr} mitigate incomplete modalities through prompt learning. Some recovery-based methods
\cite{recover5}, \cite{recover1}, \cite{MM24robustSenDistribution}, \cite{MM24robustSenMOE},  \cite{recover4} have been proven to be effective in robust multimodal learning. Furthermore, researchers in the relevant field have proposed methods to enhance the robustness of multimodal learning. \cite{cai2025multi} proposed a causal-guided adaptive multimodal diffusion network to address the displacement phenomenon that may occur when samples from different source domains interfere with each other during the learning process. NORM-TR \cite{liu2025noise} reduces the impact of noise on multimodal learning through a noise perception learning method.

\begin{figure*}[h]
  
  \centering
  \includegraphics[width=\linewidth]{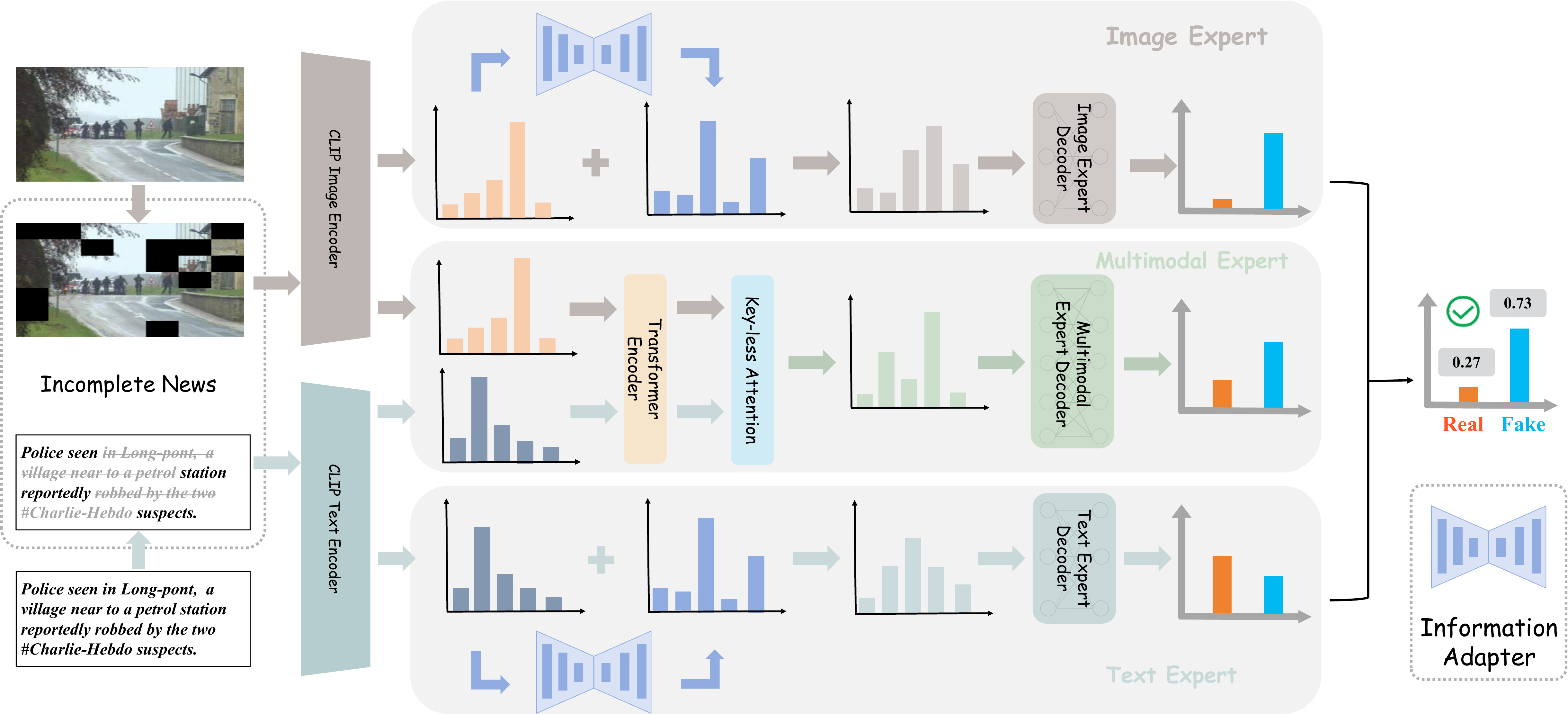}
  \caption{Overview of MMLNet framework.}
  \label{fig:framework}
\end{figure*}

However, these modality recovery-based methods are not well-suited for MMR, as they inherently rely on modality consistency. Misinformation is often embedded in modality inconsistencies, making modality recovery detrimental to the model's performance. Therefore, we utilize the modal feature distribution learned from multiple perspectives by MMLNet, make up for the missing features with the invisible feature distribution through the incomplete modality adapter, and effectively enhance the robustness of the modal features through modality missing learning.

\section{Method}
\subsection{Problem Definition}
\subsubsection{Multimodal Misinformation recognition}
The MMR task is a typical binary classification problem, where the positive class corresponds to \textit{fake news} and the negative class corresponds to \textit{real news}.
Given a multi-modal news dataset \( \mathcal{D} \), where 
\noindent \( \mathcal{D} = \{x_i,y_{i} \mid x_i = (x_{i}^{T}, x_{i}^{V})\}_{i=1}^{N}. \) 
\noindent \( x_{i}^{T}\)represents textual information, \noindent \( x_{i}^{V}\)represents visual information, and \noindent \( y_{i}\) denotes the corresponding label. To effectively distinguish fake news, the model's objective is to learn the probability distribution of \( \mathbb{P}(y \mid x) \).

\subsubsection{Incomplete Modality}
We focus on fine-grained modality incompleteness by preprocessing the raw data to obtain incomplete modality data. The text modality exhibits incompleteness at the word-level, while the image modality is incomplete at the \( n \times n \) patch-level. Specifically, given a news text \( x^{T} = \{t_1, t_2, \dots, t_m\} \), where \( t_i \) represents the \( i \)-th word and \( m \) denotes the sentence length. We introduce text modality incompleteness by randomly removing \( M^{T}\% \) of the words. The remaining text contains \( \hat{m} = m \times (1 - M^{T}\%) \) words, represented as \( \hat{x}^{T} = \{t_1, t_2, \dots, t_{\hat{m}}\} \). The corresponding news image is represented as \( x^{V} = \{v_1, v_2, \dots, v_{n \times n}\} \), where the image is divided into \( n \times n \) patches, and \( v_i \) represents the \( i \)-th patch. Under an image missing rate of \( M^{V}\% \), we randomly mask \( M^{V}\% \) of the image patches, leaving \( \hat{n} = n \times n \times (1 - M^{V}\%) \) valid patches, denoted as \( \hat{x}^{V} = \{v_1, v_2, \dots, v_{\hat{n}}\} \).  

Thus, the newly constructed incomplete multimodal news data is represented as \( \hat{x} = \{\hat{x}^{T}, \hat{x}^{V}\} \).

Formally, under the incomplete modality setting, our new optimization objective is formulated using maximum likelihood estimation, expressed as follows:
\begin{align}
\max \mathcal{O}=\mathbb{E}_{\left(\hat{x}, y\right) \sim p_{\text {Train }}} \log p\left(y \mid \hat{x}\right)
\end{align}
where \( p_{\text{train}} \) represents the distribution of incomplete modalities in the training data.

\subsection{Overview Framework}
As shown in Figure \ref{fig:framework}, the overall architecture of our proposed MMLNet consists of three key components designed for incomplete modalities: a Modality-specific Multi-expert Reasoning  paradigm, an Incomplete Modality Adapter module, and a Modality missing learning module.

\subsection{Multi-Expert Reasoning Module}
\subsubsection{Expert of Text-Modality}
In multimodal learning, the text modality typically plays a dominant role due to its higher information density. An incomplete text modality significantly impacts model performance; therefore, modeling text modality experts is crucial for the MMR task.

With the multimodal input, we adopt the CLIP model (Contrastive Language-Image Pretraining) , a vision-and-language pre-trained model that has demonstrated significant performance across various vision-and-language downstream tasks , to extract features of the textual modality and the image modality.

Given a news text \( x^{T} = \{t_1, t_2, \dots, t_m\} \), where \( t_i \) represents the \( i \)-th word and \( m \) denotes the sentence length. The CLIP text encoder \(\mathbb{T}\) outputs the corresponding representation \(H\):

\begin{align}
H = \mathbb{T}(x^{T}) = ({h}_\text{{[CLS]}}, {h}_{1}, {h}_{2}, ..., {h}_{m})
\end{align}

We employ ${h}_\texttt{{[CLS]}}$ to model the distribution of the text modality expert, and we employ the text modality adapter $f_A^T(\cdot)$ to enhance feature robustness, as shown below:
\begin{equation}
    F_T =f_A^T(h_\text{{[CLS]}})
\end{equation}

\begin{align}
{y}^{h}=\text{softmax}(\boldsymbol{W}F_T+ \boldsymbol{b})
\end{align}
where $\boldsymbol{y}^{h}$ is the output distribution of text modality expert.

\subsubsection{Expert of Image-Modality}
The image modality is typically a complementary modality to text, and its integration with the text modality facilitates the assessment of news authenticity. We employ the CLIP image encoder to model the image modality expert.

For the corresponding image \( x^{V} = \{v_1, v_2, \dots, v_{n}\} \), the CLIP image encoder \(\mathbb{V}\) outputs the corresponding representation \(R\):

\begin{align}
R = \mathbb{V}(x^{V}) = ({r}_\text{{[CLS]}}, {r}_{1}, {r}_{2}, ..., {r}_{n})
\end{align}
where \(n\) represents the number of image patches.

Similarly, $\mathbf{r}_\text{{[CLS]}}$ is used to model the image modality expert, and we employ the image modality adapter $f_A^I(\cdot)$ to enhance feature robustness, as shown below:

\begin{equation}
    F_I =f_A^I({r}_\text{{[CLS]}})
\end{equation}

\begin{align}
{y}^{r}=\text{softmax}(\boldsymbol{W}F_I+ \boldsymbol{b})
\end{align}

where ${y}^{v}$ is the output distribution of image modality expert.

\begin{figure*}[h]
  
  \centering
  \includegraphics[width=\linewidth]{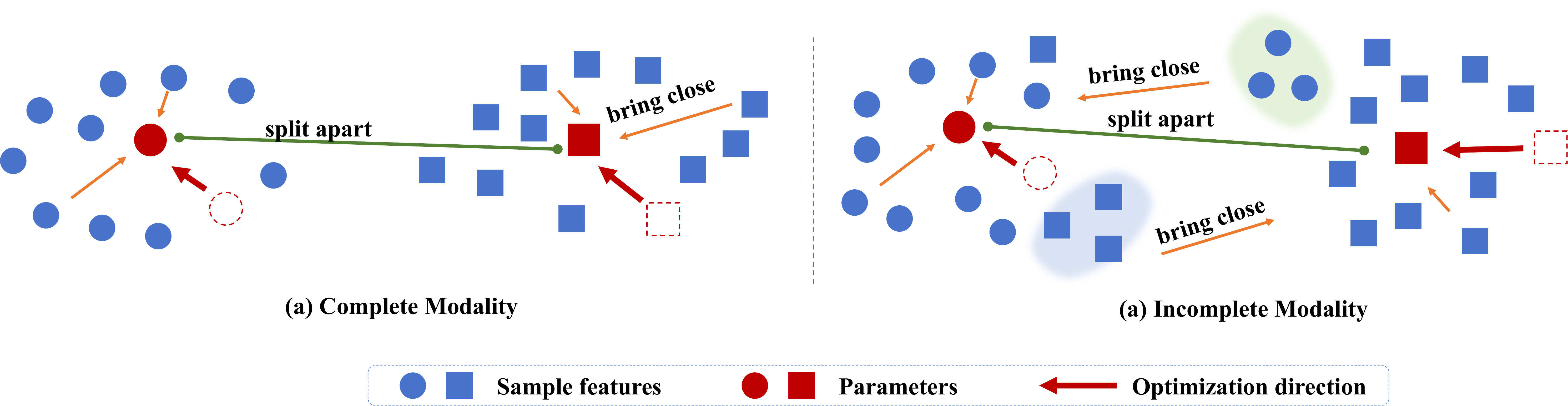}
  \caption{Distribution of features in space for complete and incomplete modality.}
  \label{fig:figcl}
\end{figure*}

\subsubsection{Expert of Multi-Modality}
Modeling the distribution between text modality and image modality is advantageous for capturing deep-level deceptive information. However, incomplete modality information can mislead the modeling of inter-modality distributions, thereby hindering the recognition of deceptive information. In this work, we employ a transformer encoder to model the joint expert distribution across multiple modalities.

Specifically, given the yielded text representation \( H \) and image representation \( R \), we first concat them as follows: 
\begin{align}
F = (H, R) = (h_{\text{[CLS]}}, h_1, \dots, h_m, r_{\text{[CLS]}}, r_1, \dots, r_n)
\end{align}
Then, we apply different linear functions to obtain the corresponding queries \( Q \), keys \( K \) and values \( V \), and the updated representation \( \hat{F} \) can be denoted as
\begin{align}
\hat{F} = (\hat{h}_{\text{[CLS]}},\hat{h}_1, \dots, \hat{h}_m,\hat{r}_{\text{[CLS]}}, \hat{r}_1, \dots, \hat{r}_n)=\notag\\\text{softmax} \left( \frac{QK^\top}{\sqrt{d_k}} \right) V
\end{align}
where \( d_k \) is the mapped dimension.

we use attention mechanism to further fuse the image-text interaction feature \( f \) by calculating:
\begin{align}
p_t, p_v = \text{softmax} (\boldsymbol W (\hat{h}_{\text{[CLS]}}, \hat{r}_{\text{[CLS]}}) + b)
\end{align}
\begin{align}
f = p_t \hat{h}_{\text{[CLS]}} + p_v \hat{r}_{\text{[CLS]}})
\end{align}
\begin{align}
y^f = \text{softmax}(\boldsymbol Wf + \boldsymbol b)
\end{align}

where \( y^f \) is output distribution.

\subsubsection{Multi-Expert Reasoning Paradigm}
Given the obtained expert feature distributions \( y^h, y^r, y^f \), We employ a dynamic routing network to balance the weights of different expert distributions, aiming to better integrate the judgments of various experts in determining the authenticity of news.

\begin{align}
{y}^{o} = \sum_{M\in \left \{ h,r,f \right \} } {\lambda }_{o}^{M} {y}^{M},{\lambda }_{o} \in\left [ 0,1 \right ] 
\end{align}

Here, \( \lambda_o \) is a learnable parameter, \( y^o \) can be regarded as harnessing rich features from the text modality $h$, image modality $r$, and image-text interaction modality $f$, which enables MMLNet to achieve better robustness in the presence of missing modality information.

\subsection{Incomplete Modality Adapters}
Mainstream MMR tasks typically employ models pre-trained on large-scale datasets as encoders, which may lead to overfitting of feature distributions, a problem that becomes more pronounced when modality information is missing. Inspired by \cite{clip_ada}, we introduce lightweight feature adapters for single-modality image and text, enabling the learning of new modality features to compensate for the missing information while preserving the original distributions learned by the pre-trained models.

We utilize residual connections to merge the feature distributions of images and text with the new distributions from adapters, enabling the model to balance compensated modality features and disentangled image and text features. With the image and text Feature Adapters $A^I(\cdot)$ and $A^T(\cdot)$ respectively, the final unimodal representations $F_I, F_T$ are obtained as follows:
\begin{equation}
    F_I = \alpha A^I({r}_\text{{[CLS]}}) + (1 - \alpha) {r}_\text{{[CLS]}} =f_A^I({r}_\text{{[CLS]}})
\end{equation}
\begin{equation}
    F_T = \alpha A^T({h}_\text{{[CLS]}}) + (1 - \alpha){h}_\text{{[CLS]}}=f_A^T({h}_\text{{[CLS]}})
\end{equation}
Where $\alpha$ is residual ratio to maintain harmony between these two modules. 
We do not focus on compensating for multimodal feature distributions; instead, we aim to extract misinformation from interactions within the original single modality features.

\subsection{Modality Missing Learning}
As shown in Figure \ref{fig:figcl}, intuitively, when samples with opposite labels show greater semantic similarity, it is more difficult for the model to distinguish the authenticity of the news. On the contrary, when there is a significant difference in the semantic similarity of samples with the same label, the model tends to make inconsistent inferences about their authenticity. Missing modality information worsens this phenomenon.

To enhance the performance of MMLNet in MMR tasks under conditions of incomplete modality information, we propose a modality missing learning algorithm based on a vanilla multimodal contrastive learning framework. This simple yet effective approach aims to mitigate the negative effects of missing modality information by constraining the feature distributions across multiple modalities. 
Specifically, for each training sample \(x\), we first define the corresponding positive sample set \(\mathcal{S}_p\) and negative sample set \(\mathcal{S}_n\) for it. Here, \(\mathcal{S}_p\) contains samples with the same label as \(x\), while \(\mathcal{S}_n\) contains samples with the opposite label as \(x\). As shown in Figure \ref{fig:figcl}, we aim to learn a robust latent representation of \(x\) by attracting \(x\) to the representation of samples in \(\mathcal{S}_p\) and repidating it away from the representation of samples in \(\mathcal{S}_n\) as follows:

\begin{align}
\mathcal{L}_{m} = \frac{1}{\left|\mathcal{S}_{p}\right|} \sum_{p \in \mathcal{S}_{p}}-\log \frac{\exp \left(f(\mathbf{h}) \cdot f\left(\mathbf{h}_{p}\right) / \tau\right)}{\sum_{n \in \mathcal{S}_{n}} \exp \left(f(\mathbf{h}) \cdot f\left(\mathbf{h}_{n}\right) / \tau\right)}
\end{align}

where $f(\cdot)$ denotes an MLP projection. \(\tau\) is a temperature coefficient.  \(h_p\) and \(h_n\) are the final representations of the positive sample \(p\) and negative sample \(n\).

However, in incomplete modal MMR task, it is suboptimal to weight all positive and negative samples equally. This alignment objective is not fully compatible with multimodal misinformation recognition with incomplete modalities. Based on the aforementioned perceptions, our model encounters two common but difficult problems caused by the lack of modality information: 1) distinguishing samples with similar semantic information but opposite labels due to incomplete modal information, and 2) narrowing the gap between samples with different semantic information but the same label due to incomplete modal information. To accomplish this, we introduce a label-aware adaptive weighting strategy. Specifically, we measure the semantic similarity of two samples by measuring the cosine similarity,  each sample is re-weighted as follows:

\begin{align}
w_{c} = \left\{\begin{array}{l}
1-cos(h_c,h),\left(c \in \mathcal{S}_{p}\right) \\
1+cos(h_c,h),\left(c \in \mathcal{S}_{n}\right)
\end{array}\right.
\end{align}

where \( w_c \) is the label-aware weight for sample \( c \). \( h_c \) is the representation vector of sample \( c \). Ultimately, the contrastive loss is refined as follows:

\begin{align}
\hat{\mathcal{L}}_{m} = \frac{1}{\left|\mathcal{S}_{p}\right|} \sum_{p \in \mathcal{S}_{p}}-\log \frac{w_{p} \cdot \exp \left(f(\mathbf{h}) \cdot f\left(\mathbf{h}_{p}\right) / \tau\right)}{\sum_{n \in \mathcal{S}_{n}} w_{n} \cdot \exp \left(f(\mathbf{h}) \cdot f\left(\mathbf{h}_{n}\right) / \tau\right)}
\end{align}

where \( w_p \) and \( w_n \) denote the weights assigned for a positive sample \( p \) and a negative sample \( n \), respectively.

In fact, modality missing learning is not only effective for multimodal features. We also apply it to text modality features $F_T$ and image modality features $F_I$ to further constrain the distribution of unimodal features and enhance the robustness of the MMLNet model in the incomplete modality MMR task.

\subsection{Overall Optimization}

We employ the cross-entropy loss to optimize as follows,
\begin{align}
\mathcal{L}_{c}=-\sum_{i=1}^{N} y_{i}^{T}~log~\hat{y}_{i}
\end{align}

Here, $y$ represents the ground truth, $\hat{y}$ stands for the probability of the predicted label corresponding to the i-th image-text pair.

Finally, we combine the classification and modality missing learning loss functions together to optimize the whole model as follows,

\begin{align}
\mathcal{L}=\sum_{M \in\{I, T, F\}} \lambda_{o}^{M} \left (\lambda_{c}  \mathcal{L}_{c}^{M}+\lambda_{m}  \mathcal{L}_{m}^{M}\right)
\end{align}

where \(\lambda_{c}\), \(\lambda_{o}\) and \(\lambda_{m}\) is a hyperparameter that controls the loss weight to balance learning. Note that the hyperparameter \(\lambda_{o}\) is consistent with Eq. 13.

\section{Experiments}

In this section, we conduct relevant experiments to address several research questions. Through these experimental results, we systematically evaluate the effectiveness of MMLNet in tackling key challenges in incomplete modality misinformation recognition:

\begin{itemize}
\item \textbf{RQ1:} Can MMLNet effectively enhance the performance of misinformation recognition?
\item \textbf{RQ2:} Is the proposed design robust for misinformation recognition under incomplete modalities?
\item \textbf{RQ3:} Do individual components contribute to improving misinformation recognition capability?
\item \textbf{RQ4:} Can modallity missing learning significantly improve incomplete modality MMR?
\item \textbf{RQ5:} How do MMLNet accurately assess the authenticity of the news?
\item \textbf{RQ6:} Does MMLNet have limitations in determining the authenticity of news with incomplete information?
\end{itemize}

\subsection{Datasets}
We evaluate the performance of MMLNet on three real-world datasets: Weibo \cite{weibo}, Weibo21 \cite{weibo21}, and Pheme \cite{pheme}, covering both Chinese and English. The Weibo dataset is the most widely used Chinese misinformation recognition dataset, containing fake and real news from Sina Weibo between 2012 and 2016, verified by authoritative sources such as Xinhua News Agency and Sina's official debunking system. The Weibo21 dataset, released in 2021, is an updated version of the Weibo dataset that includes more recent Weibo news. Pheme is an English-language dataset consisting of tweets from Twitter, focusing on five breaking news events. Each event comprises a set of posts containing a substantial amount of textual and visual content, along with corresponding annotations. Three events are randomly selected, and their associated tweets are used for training, while tweets from the remaining events are utilized for testing. For comparison with baseline methods, we follow previous studies \cite{NSLMaaai24}, \cite{MOE_WWW25},  \cite{DGAmm24}, \cite{pheme_pro} and use the same preprocessed dataset.

\begin{figure*}[htbp]

\begin{subfigure}[t]{0.192\textwidth}
\centering
\includegraphics[width=3.44cm]{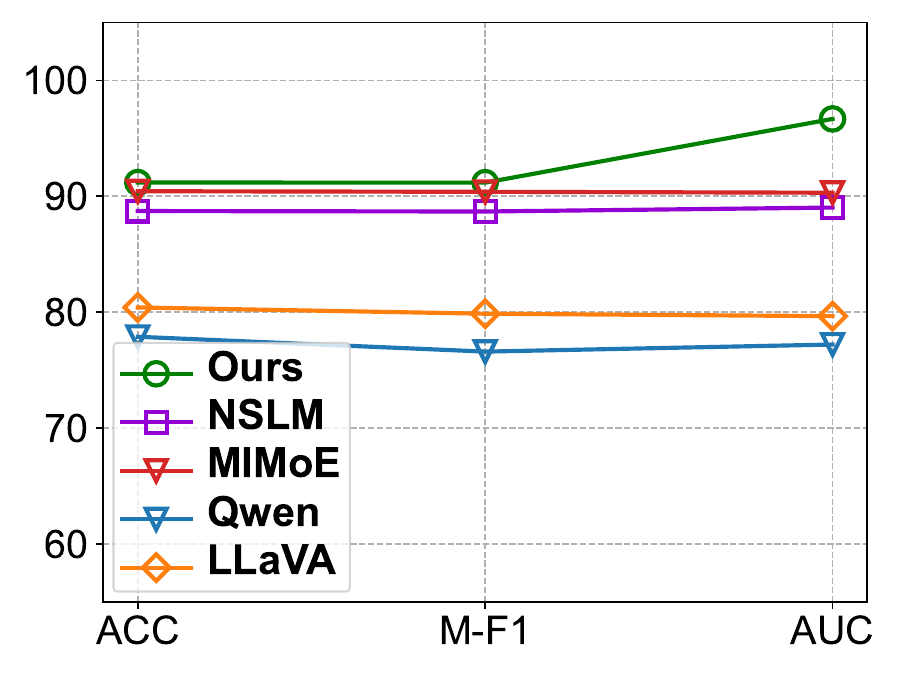}
\caption{T↓0\%, I↓100\%.}
\end{subfigure}
\begin{subfigure}[t]{0.192\textwidth}
\centering
\includegraphics[width=3.44cm]{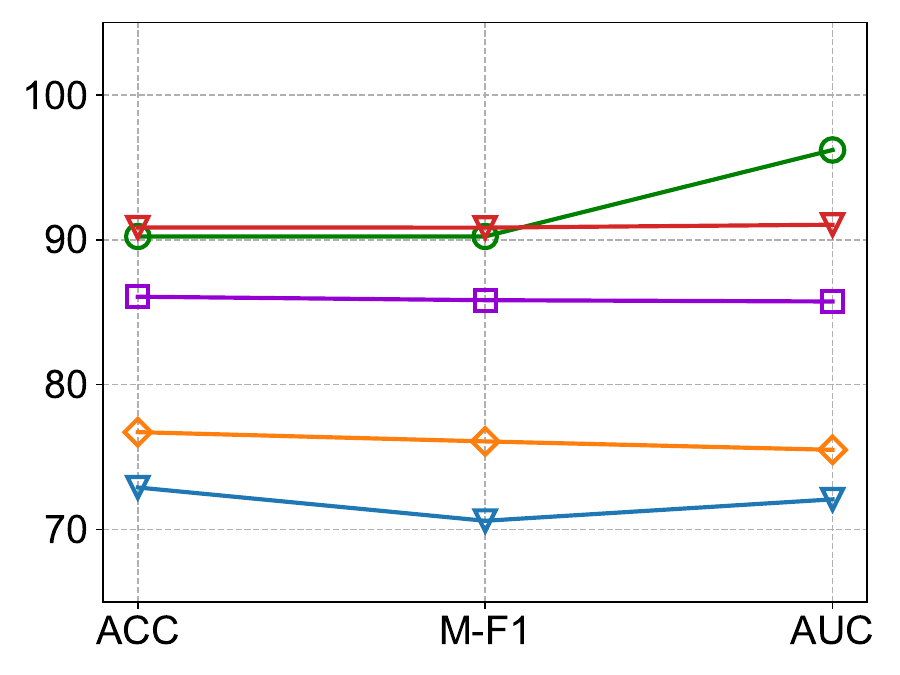}
\caption{T↓25\%, I↓75\%.}
\end{subfigure}
\centering
\begin{subfigure}[t]{0.192\textwidth}
\centering
\includegraphics[width=3.44cm]{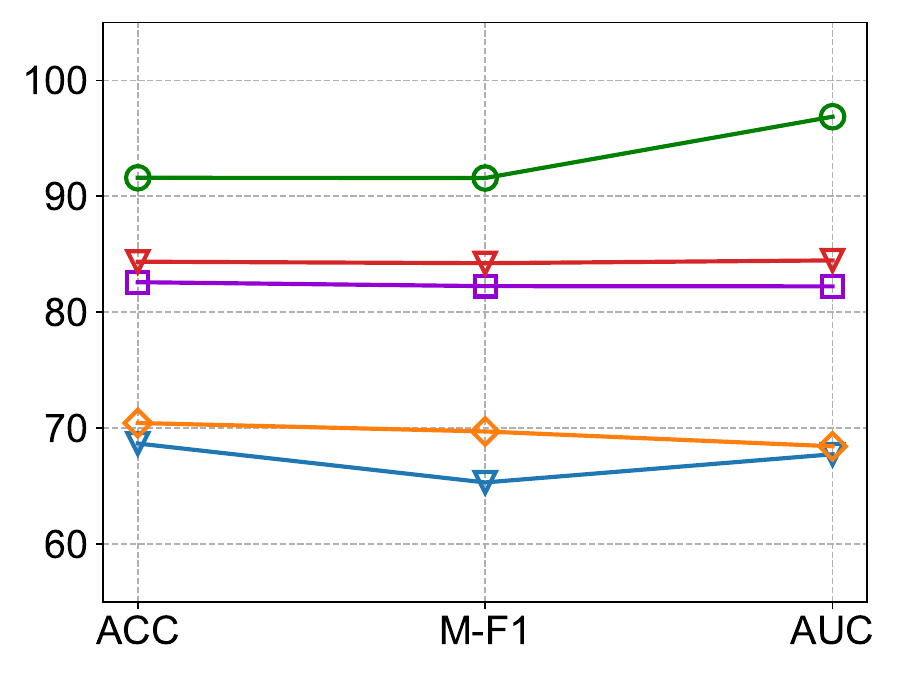}
\caption{T↓50\%, I↓50\%.}
\end{subfigure}
\begin{subfigure}[t]{0.192\textwidth}
\centering
\includegraphics[width=3.44cm]{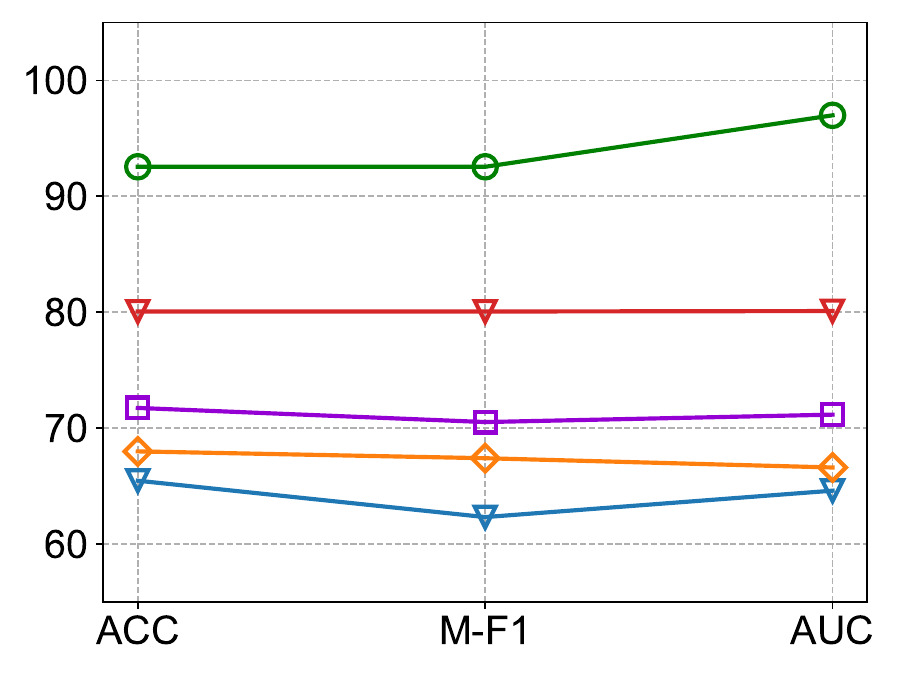}
\caption{T↓75\%, I↓25\%.}
\end{subfigure}
\begin{subfigure}[t]{0.192\textwidth}
\centering
\includegraphics[width=3.44cm]{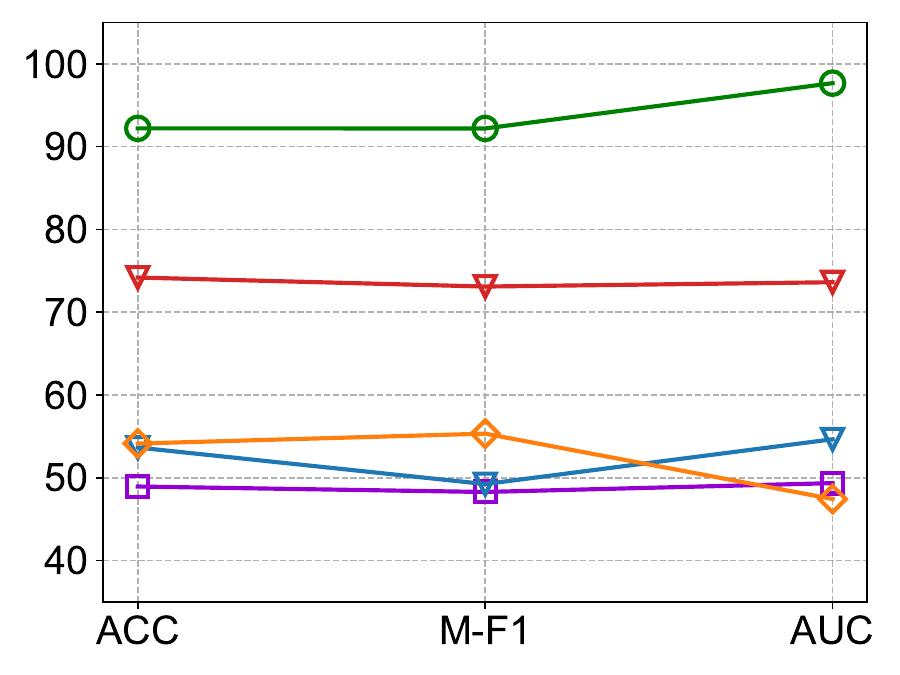}
\caption{T↓100\%, I↓0\%.}
\end{subfigure}
\caption{Experimental results of Weibo dataset. `T' and `I' respectively represent the text and image modalities.}
\label{fig:weibo}
\vspace{-0.20cm} 
\end{figure*}

\begin{figure*}[htbp]

\begin{subfigure}[t]{0.192\textwidth}
\centering
\includegraphics[width=3.44cm]{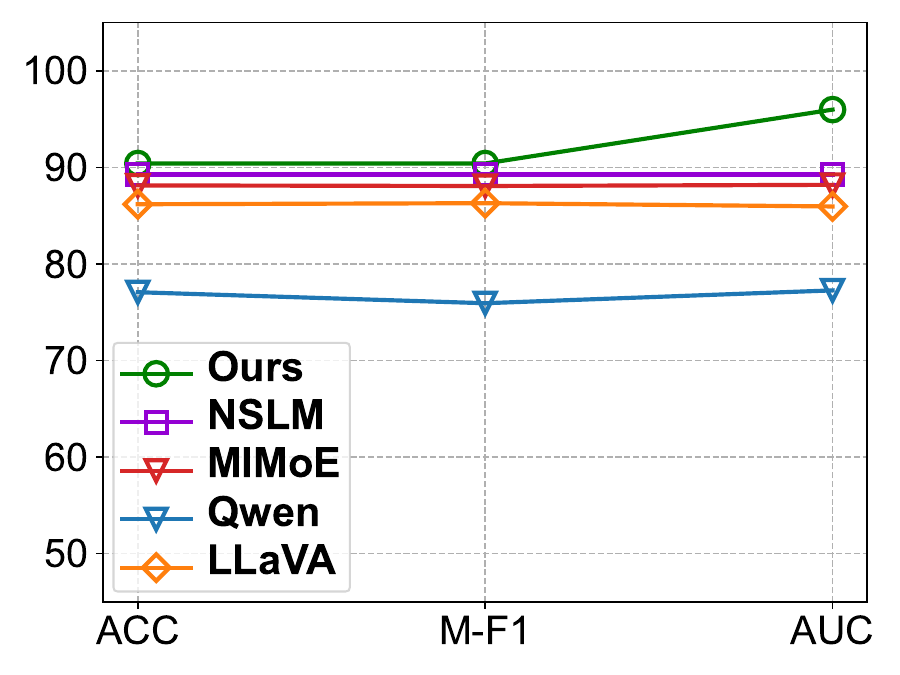}
\caption{T↓0\%, I↓100\%.}
\end{subfigure}
\begin{subfigure}[t]{0.192\textwidth}
\centering
\includegraphics[width=3.44cm]{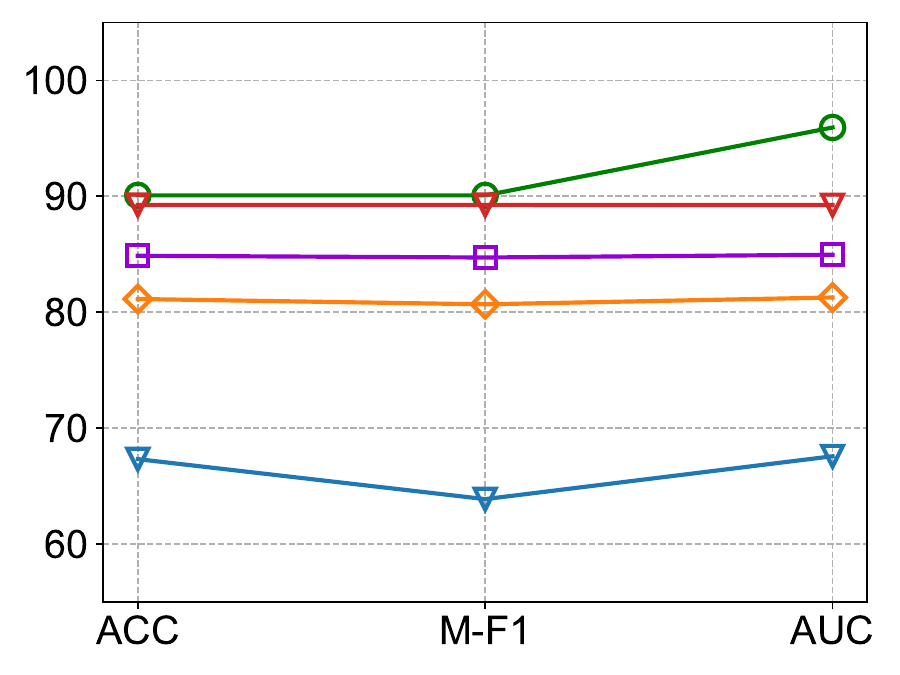}
\caption{T↓25\%, I↓75\%.}
\end{subfigure}
\centering
\begin{subfigure}[t]{0.192\textwidth}
\centering
\includegraphics[width=3.44cm]{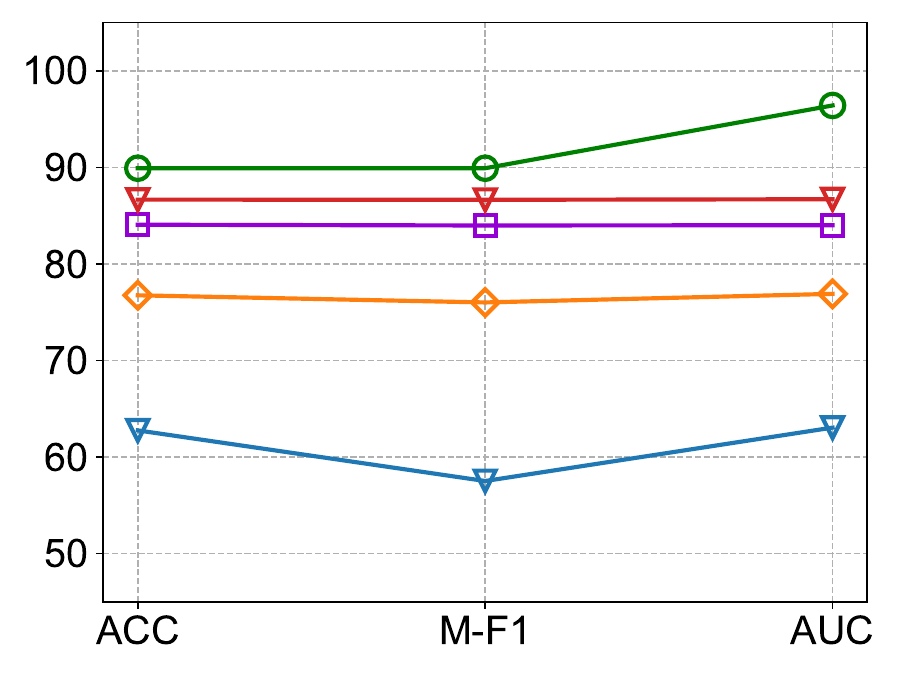}
\caption{T↓50\%, I↓50\%.}
\end{subfigure}
\begin{subfigure}[t]{0.192\textwidth}
\centering
\includegraphics[width=3.44cm]{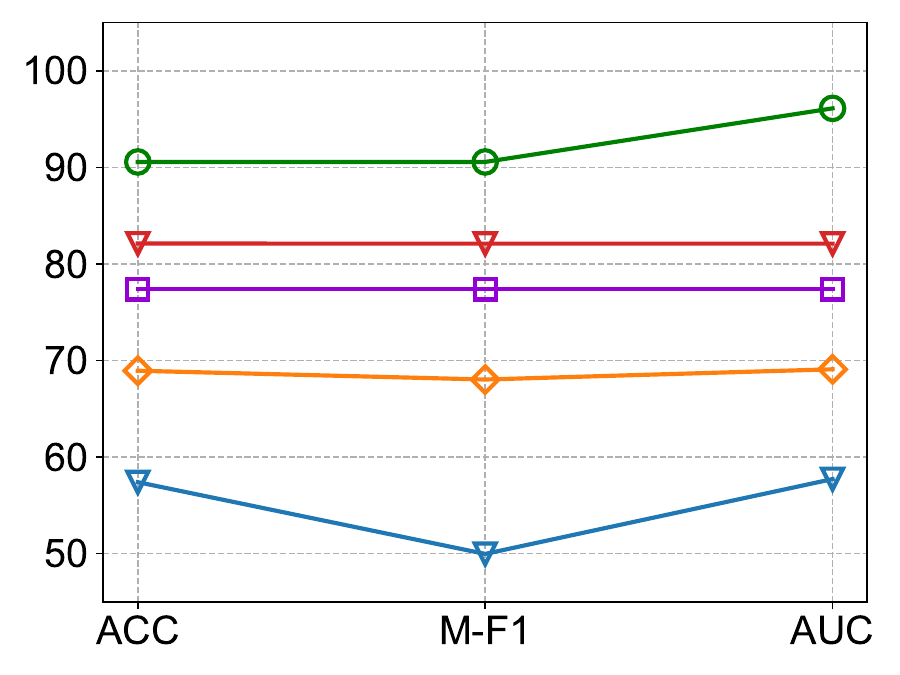}
\caption{T↓75\%, I↓25\%.}
\end{subfigure}
\begin{subfigure}[t]{0.192\textwidth}
\centering
\includegraphics[width=3.44cm]{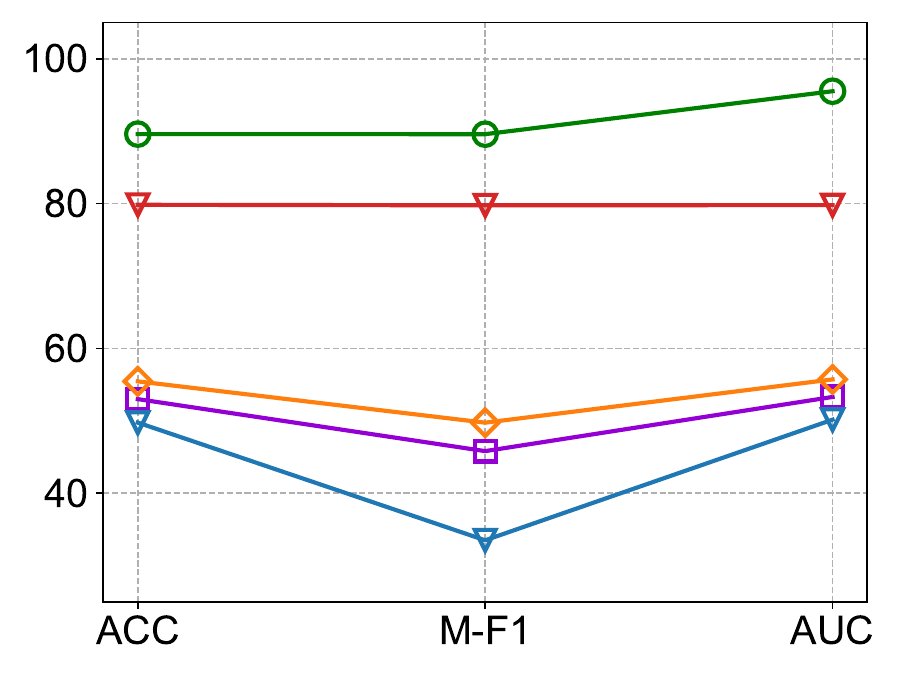}
\caption{T↓100\%, I↓0\%.}
\end{subfigure}
\caption{Experimental results of Weibo21 dataset. `T' and `I' respectively represent the text and image modalities.}
\label{fig:weibo21}
\vspace{-0.20cm} 
\end{figure*}

\begin{figure*}[htbp]

\begin{subfigure}[t]{0.192\textwidth}
\centering
\includegraphics[width=3.44cm]{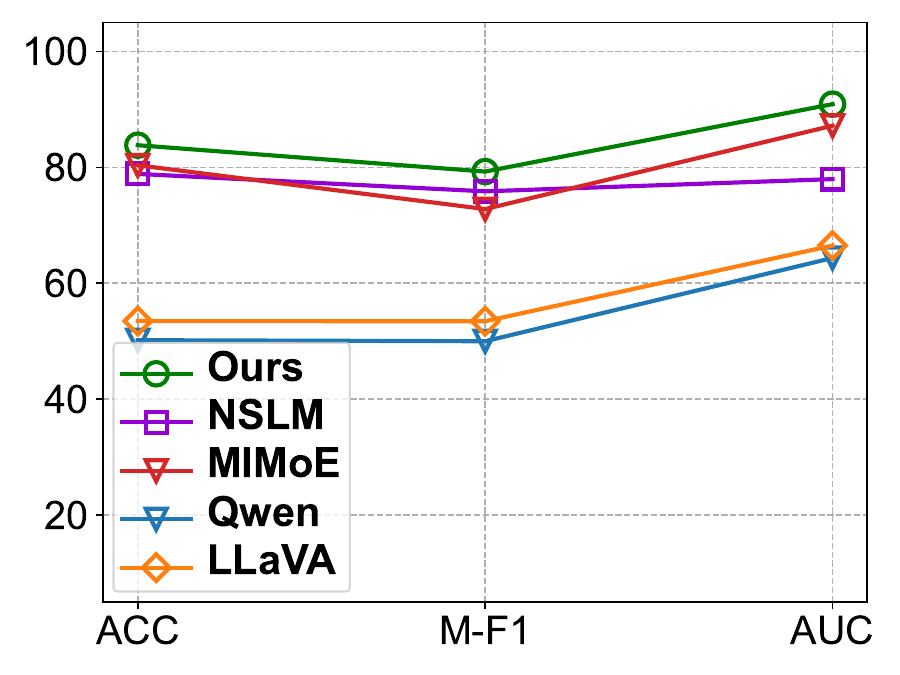}
\caption{T↓0\%, I↓100\%.}
\end{subfigure}
\begin{subfigure}[t]{0.192\textwidth}
\centering
\includegraphics[width=3.44cm]{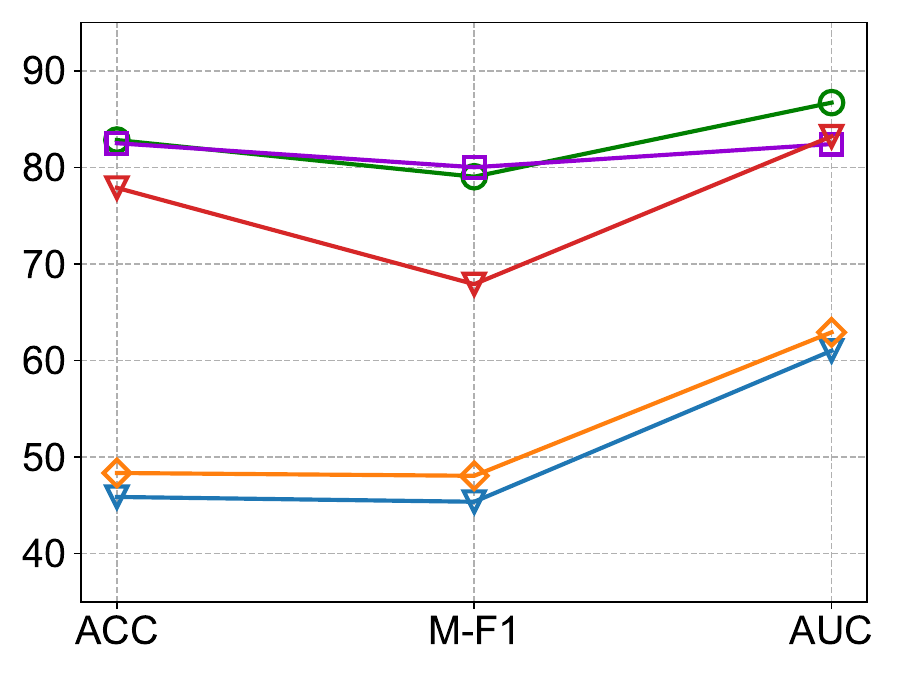}
\caption{T↓25\%, I↓75\%.}
\end{subfigure}
\centering
\begin{subfigure}[t]{0.192\textwidth}
\centering
\includegraphics[width=3.44cm]{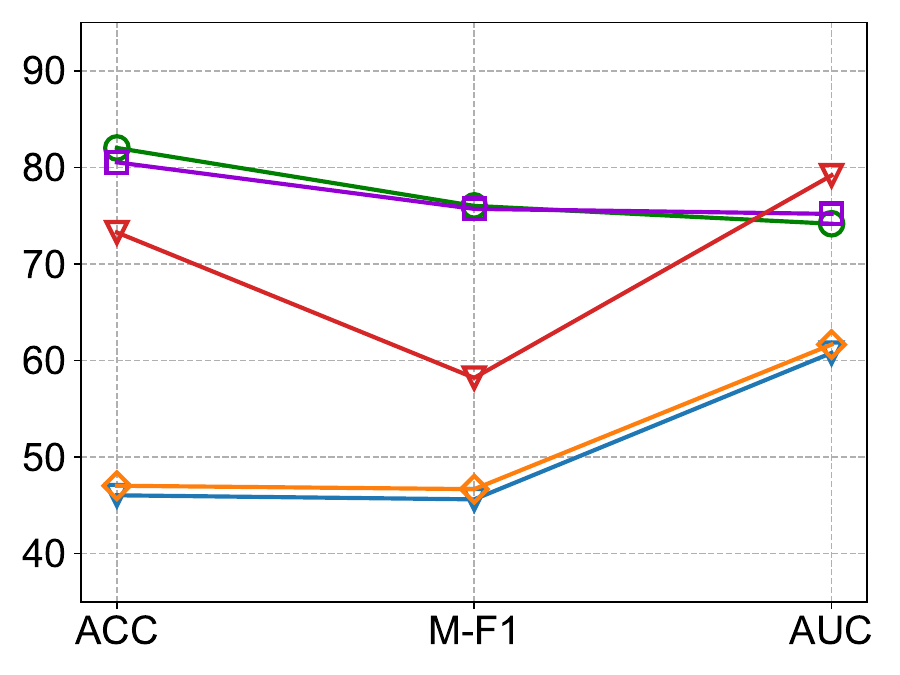}
\caption{T↓50\%, I↓50\%.}
\end{subfigure}
\begin{subfigure}[t]{0.192\textwidth}
\centering
\includegraphics[width=3.44cm]{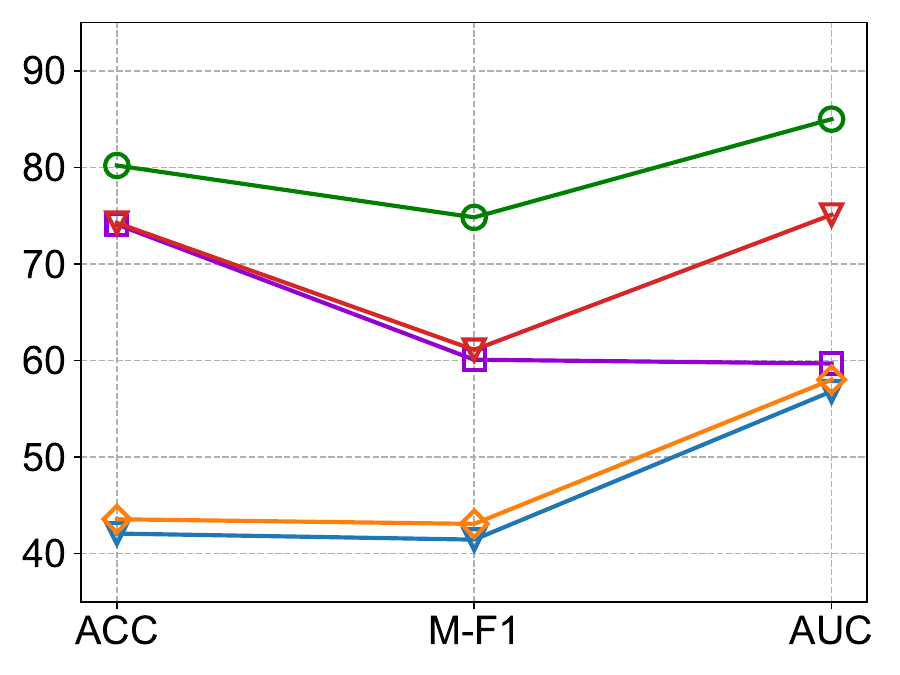}
\caption{T↓75\%, I↓25\%.}
\end{subfigure}
\begin{subfigure}[t]{0.192\textwidth}
\centering
\includegraphics[width=3.44cm]{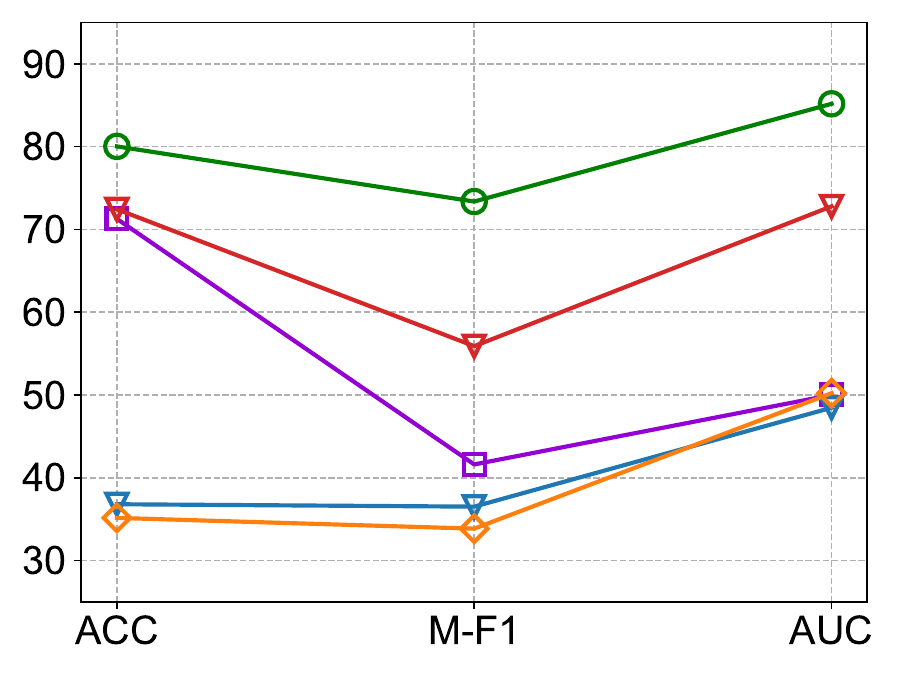}
\caption{T↓100\%, I↓0\%.}
\end{subfigure}
\caption{Experimental results of Pheme dataset. `T' and `I' respectively represent the text and image modalities.}
\label{fig:pheme}
\vspace{-0.20cm} 
\end{figure*}

\subsection{Evaluation Metrics}
The multimodal misinformation recognition problem is a binary classification task. Based on previous work \cite{NSLMaaai24}, \cite{fkaowl}, \cite{MOE_WWW25},  we apply the accuracy score (ACC), the area under the receiver operating characteristic curve (AUC), and the macro-F1 score (M-F1) as evaluation metrics.

\subsection{Baseline}
We compare the two open-sourced sota MFDN methods including NSLM \cite{NSLMaaai24} and MIMoE \cite{MOE_WWW25}. We also fine-tunes the parameters of MLLM to show the ability of MLLM in MMR in complete and incomplete modal scenarios, specifically using the open-sourced MLLM Qwen-VL-7B\footnote{https://github.com/QwenLM/Qwen-VL} and LLaVA-v1.5-13B\footnote{https://huggingface.co/liuhaotian/llava-v1.5-13b}.



\subsection{Implementation Details}

To facilitate future research on incomplete modality MMR, for handling missing text modality, we set a random seed and pre-store the missing text. For missing image modality, we store the indices of the missing patches. Specifically, the image is resized to $224\times224$ and decompose it into patches of size $32\times32$. For the simplicity of the experiment, the training set and the test set have the same missing settings.

To ensure fairness and reproducibility, MMLNet adopts a consistent set of hyperparameters across all datasets, with a random seed of 42. For simplicity, we use chinese-clip-vit-large-patch14\footnote{https://huggingface.co/OFA-Sys/chinese-clip-vit-large-patch14} as the encoder for Weibo and Weibo21, clip-vit-large-patch14\footnote{https://huggingface.co/openai/clip-vit-large-patch14} as the encoder for Pheme dataset. We implement our model using PyTorch and optimize it with AdamW (learning rate is 1e-4, CLIP  learning rate is 3e-6). We set the weight decay to 0.005, train for 15 epochs with a batch size of 16, and apply a dropout rate of 0.3. The hyperparameters are tuned based on the validation set of the Weibo dataset. Experiments were conducted on a single 24GB 4090 GPU. Code is publicly available for detailed review.

\definecolor{myorange}{RGB}{238,105,17}
\begin{table*}
\centering
  \caption{Ablation experiments on complete modalities on the Weibo and Pheme datasets. `w/o' is anabbreviation for `with', and `w/o' is anabbreviation for `without'.}
  \label{tab:ab}
\begin{tabular}{l|llllll}
\hline
\multirow{3}{*}{Method} & \multicolumn{6}{c}{Complete Modality}                     \\ \cline{2-7} 
                        & \multicolumn{3}{c}{Weibo}    & \multicolumn{3}{c}{Pheme}  \\ \cline{2-7} 
                        & ACC(\%) & F1(\%) & AUC(\%) & ACC(\%) & F1(\%) & AUC(\%) \\ \hline
NSLM                    & 92.28   & 92.28  & 92.32   & 84.65   & 81.82  & 82.88   \\
w/ ${L}_{m}$            & 93.44\textcolor{myorange}{\scriptsize↑1.16}   & 93.44\textcolor{myorange}{\scriptsize↑1.16}  & 93.54\textcolor{myorange}{\scriptsize↑1.22}   & 86.96\textcolor{myorange}{\scriptsize↑2.31}   & 84.31\textcolor{myorange}{\scriptsize↑2.49}  & 84.84\textcolor{myorange}{\scriptsize↑1.96}   \\
w/ $A$                  & 92.90\textcolor{myorange}{\scriptsize↑0.62}   & 92.89\textcolor{myorange}{\scriptsize↑0.61}  & 92.90\textcolor{myorange}{\scriptsize↑0.58}   & 85.64\textcolor{myorange}{\scriptsize↑0.99}   & 83.20\textcolor{myorange}{\scriptsize↑1.38}  & 84.78\textcolor{myorange}{\scriptsize↑1.90}   \\ \hline
MIMOE                   & 92.49   & 92.47  & 92.42   & 85.64   & 81.90  & 89.40   \\
w/ ${L}_{m}$            & 93.85\textcolor{myorange}{\scriptsize↑1.36}   & 93.85\textcolor{myorange}{\scriptsize↑1.38}  & 93.91\textcolor{myorange}{\scriptsize↑1.49}   & 87.12\textcolor{myorange}{\scriptsize↑1.48}   & 84.33\textcolor{myorange}{\scriptsize↑2.43}  & 84.45\textcolor{myorange}{\scriptsize↓4.95}   \\
w/ $A$                  & 93.37\textcolor{myorange}{\scriptsize↑0.88}   & 93.37\textcolor{myorange}{\scriptsize↑0.90}  & 93.48\textcolor{myorange}{\scriptsize↑1.06}   & 86.13\textcolor{myorange}{\scriptsize↑0.49}   & 82.20\textcolor{myorange}{\scriptsize↑0.30}  & 88.99\textcolor{myorange}{\scriptsize↓0.41}   \\ \hline
MMLNet-Ours             & 95.22   & 95.21  & 98.93   & 87.78   & 84.35  & 90.97   \\
w/o ${L}_{c}^{h}$       & 94.67\textcolor{mydarkgreen}{\scriptsize↓0.55}   & 94.65\textcolor{mydarkgreen}{\scriptsize↓0.56}  & 98.79\textcolor{mydarkgreen}{\scriptsize↓0.14}   & 87.45\textcolor{mydarkgreen}{\scriptsize↓0.33}   & 84.29\textcolor{mydarkgreen}{\scriptsize↓0.06}  & 89.53\textcolor{mydarkgreen}{\scriptsize↓1.44}   \\
w/o ${L}_{c}^{r}$       & 94.88\textcolor{mydarkgreen}{\scriptsize↓0.34}   & 94.87\textcolor{mydarkgreen}{\scriptsize↓0.34}  & 98.91\textcolor{mydarkgreen}{\scriptsize↓0.02}   & 87.29\textcolor{mydarkgreen}{\scriptsize↓0.49}   & 84.05\textcolor{mydarkgreen}{\scriptsize↓0.30}  & 89.49\textcolor{mydarkgreen}{\scriptsize↓1.48}   \\
w/o ${L}_{c}^{f}$       & 91.80\textcolor{mydarkgreen}{\scriptsize↓3.42}   & 91.80\textcolor{mydarkgreen}{\scriptsize↓3.41}  & 97.57\textcolor{mydarkgreen}{\scriptsize↓1.36}   & 86.46\textcolor{mydarkgreen}{\scriptsize↓1.32}   & 82.86\textcolor{mydarkgreen}{\scriptsize↓1.49}  & 89.60\textcolor{mydarkgreen}{\scriptsize↓1.37}   \\
w/o ${L}_{m}^{h}$       & 95.01\textcolor{mydarkgreen}{\scriptsize↓0.21}   & 95.01\textcolor{mydarkgreen}{\scriptsize↓0.20}  & 98.95\textcolor{mydarkgreen}{\scriptsize↑0.02}   & 87.12\textcolor{mydarkgreen}{\scriptsize↓0.66}   & 83.94\textcolor{mydarkgreen}{\scriptsize↓0.41}  & 90.69\textcolor{mydarkgreen}{\scriptsize↓0.28}   \\
w/o ${L}_{m}^{r}$       & 94.94\textcolor{mydarkgreen}{\scriptsize↓0.28}   & 94.94\textcolor{mydarkgreen}{\scriptsize↓0.27}  & 98.97\textcolor{mydarkgreen}{\scriptsize↑0.04}   & 86.96\textcolor{mydarkgreen}{\scriptsize↓0.82}   & 83.93\textcolor{mydarkgreen}{\scriptsize↓0.42}  & 89.64\textcolor{mydarkgreen}{\scriptsize↓1.33}   \\
w/o ${L}_{m}^{f}$       & 93.31\textcolor{mydarkgreen}{\scriptsize↓1.91}   & 93.27\textcolor{mydarkgreen}{\scriptsize↓1.94}  & 98.41\textcolor{mydarkgreen}{\scriptsize↓0.52}   & 86.30\textcolor{mydarkgreen}{\scriptsize↓1.48}   & 82.75\textcolor{mydarkgreen}{\scriptsize↓1.60}  & 89.84\textcolor{mydarkgreen}{\scriptsize↓1.13}   \\
w/o $A$                 & 94.19\textcolor{mydarkgreen}{\scriptsize↓1.03}   & 94.18\textcolor{mydarkgreen}{\scriptsize↓1.03}  & 98.53\textcolor{mydarkgreen}{\scriptsize↓0.40}   & 86.63\textcolor{mydarkgreen}{\scriptsize↓1.15}   & 83.29\textcolor{mydarkgreen}{\scriptsize↓1.06}  & 89.80\textcolor{mydarkgreen}{\scriptsize↓1.17}   \\ \hline
\end{tabular}
\end{table*}

\begin{table*}[]
\centering
  \caption{Ablation experiments on incomplete modalities on the Weibo and Pheme datasets. In the incomplete modality setting, the text modality is missing 50\% and the image modality is missing 50\%. `w/o' is anabbreviation for `with', and `w/o' is anabbreviation for `without'.}
  \label{tab:ab}
\begin{tabular}{l|llllll}
\hline
\multirow{3}{*}{Method} & \multicolumn{6}{c}{Incomplete Modality}                     \\ \cline{2-7} 
                        & \multicolumn{3}{c}{Weibo}    & \multicolumn{3}{c}{Pheme}  \\ \cline{2-7} 
                        & ACC(\%) & F1(\%) & AUC(\%) & ACC(\%) & F1(\%) & AUC(\%) \\ \hline
NSLM                    & 82.59   & 82.25  & 82.23   & 78.87   & 75.86  & 77.97   \\
w/ ${L}_{m}$            & 86.68\textcolor{myorange}{\scriptsize↑4.09}   & 86.68\textcolor{myorange}{\scriptsize↑4.43}  & 86.78\textcolor{myorange}{\scriptsize↑4.55}   & 81.18\textcolor{myorange}{\scriptsize↑2.31}   & 76.44\textcolor{myorange}{\scriptsize↑0.58}  & 75.82\textcolor{myorange}{\scriptsize↓2.15}   \\
w/ $A$                  & 85.87\textcolor{myorange}{\scriptsize↑3.28}   & 85.84\textcolor{myorange}{\scriptsize↑3.59}  & 86.10\textcolor{myorange}{\scriptsize↑3.87}   & 80.52\textcolor{myorange}{\scriptsize↑1.65}   & 75.70\textcolor{myorange}{\scriptsize↓0.16}  & 75.18\textcolor{myorange}{\scriptsize↓2.79}   \\ \hline
MIMOE                   & 84.36   & 84.23  & 84.74   & 73.26   & 58.20  & 79.17   \\
w/ ${L}_{m}$            & 89.48\textcolor{myorange}{\scriptsize↑5.12}   & 89.47\textcolor{myorange}{\scriptsize↑5.24}  & 89.47\textcolor{myorange}{\scriptsize↑4.73}   & 75.08\textcolor{myorange}{\scriptsize↑1.82}   & 65.90\textcolor{myorange}{\scriptsize↑7.70}  & 78.99\textcolor{myorange}{\scriptsize↓0.18}   \\
w/ $A$                  & 86.89\textcolor{myorange}{\scriptsize↑2.53}   & 86.74\textcolor{myorange}{\scriptsize↑2.51}  & 86.64\textcolor{myorange}{\scriptsize↑1.9}   & 74.91\textcolor{myorange}{\scriptsize↑1.65}   & 61.50\textcolor{myorange}{\scriptsize↑3.30}  & 78.47\textcolor{myorange}{\scriptsize↓0.70}   \\ \hline
MMLNet-Ours             & 91.60   & 91.58  & 96.87   & 82.01   & 76.01  & 74.16   \\
w/o ${L}_{c}^{h}$       & 90.10\textcolor{mydarkgreen}{\scriptsize↓1.50}   & 90.09\textcolor{mydarkgreen}{\scriptsize↓1.49}  & 96.41\textcolor{mydarkgreen}{\scriptsize↓0.46}   & 81.02\textcolor{mydarkgreen}{\scriptsize↓0.99}   & 74.58\textcolor{mydarkgreen}{\scriptsize↓1.43}  & 85.25\textcolor{mydarkgreen}{\scriptsize↑11.09}   \\
w/o ${L}_{c}^{r}$       & 91.33\textcolor{mydarkgreen}{\scriptsize↓0.27}   & 91.32\textcolor{mydarkgreen}{\scriptsize↓0.26}  & 97.12\textcolor{mydarkgreen}{\scriptsize↑0.25}   & 81.18\textcolor{mydarkgreen}{\scriptsize↓0.83}   & 73.12\textcolor{mydarkgreen}{\scriptsize↓2.89}  & 70.67\textcolor{mydarkgreen}{\scriptsize↑3.49}   \\
w/o ${L}_{c}^{f}$       & 89.21\textcolor{mydarkgreen}{\scriptsize↓2.39}   & 89.15\textcolor{mydarkgreen}{\scriptsize↓2.43}  & 95.83\textcolor{mydarkgreen}{\scriptsize↓1.04}   & 79.86\textcolor{mydarkgreen}{\scriptsize↓2.15}   & 71.94\textcolor{mydarkgreen}{\scriptsize↓4.07}  & 84.75\textcolor{mydarkgreen}{\scriptsize↑10.59}   \\
w/o ${L}_{m}^{h}$       & 90.92\textcolor{mydarkgreen}{\scriptsize↓0.68}   & 90.91\textcolor{mydarkgreen}{\scriptsize↓0.67}  & 96.76\textcolor{mydarkgreen}{\scriptsize↓0.11}   & 80.52\textcolor{mydarkgreen}{\scriptsize↓1.49}   & 73.98\textcolor{mydarkgreen}{\scriptsize↓2.03}  & 84.44\textcolor{mydarkgreen}{\scriptsize↑10.28}   \\
w/o ${L}_{m}^{r}$       & 91.26\textcolor{mydarkgreen}{\scriptsize↓0.34}   & 91.24\textcolor{mydarkgreen}{\scriptsize↓0.34}  & 97.06\textcolor{mydarkgreen}{\scriptsize↑0.19}   & 81.51\textcolor{mydarkgreen}{\scriptsize↓0.50}   & 74.85\textcolor{mydarkgreen}{\scriptsize↓1.16}  & 72.79\textcolor{mydarkgreen}{\scriptsize↑1.37}   \\
w/o ${L}_{m}^{f}$       & 89.76\textcolor{mydarkgreen}{\scriptsize↓1.84}   & 89.73\textcolor{mydarkgreen}{\scriptsize↓1.85}  & 96.24\textcolor{mydarkgreen}{\scriptsize↓0.63}   & 80.19\textcolor{mydarkgreen}{\scriptsize↓1.82}   & 74.82\textcolor{mydarkgreen}{\scriptsize↓1.19}  & 84.95\textcolor{mydarkgreen}{\scriptsize↑10.97}   \\
w/o $A$                 & 89.82\textcolor{mydarkgreen}{\scriptsize↓1.78}   & 89.82\textcolor{mydarkgreen}{\scriptsize↓1.76}  & 96.80\textcolor{mydarkgreen}{\scriptsize↓0.07}   & 80.69\textcolor{mydarkgreen}{\scriptsize↓1.32}   & 75.40\textcolor{mydarkgreen}{\scriptsize↓0.61}  & 84.47\textcolor{mydarkgreen}{\scriptsize↑10.31}   \\ \hline
\end{tabular}
\end{table*}

\begin{figure*}[htbp]

\begin{subfigure}[t]{0.192\textwidth}
\centering
\includegraphics[width=3.44cm]{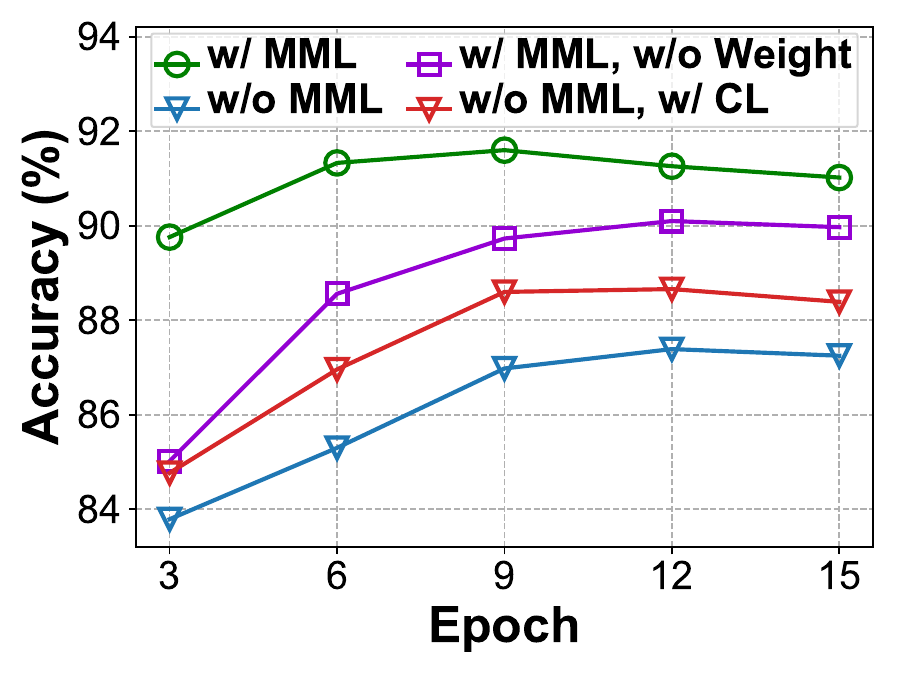}
\caption{Training Curves.}
\label{fig:curve}
\end{subfigure}
\begin{subfigure}[t]{0.192\textwidth}
\centering
\includegraphics[width=3.44cm]{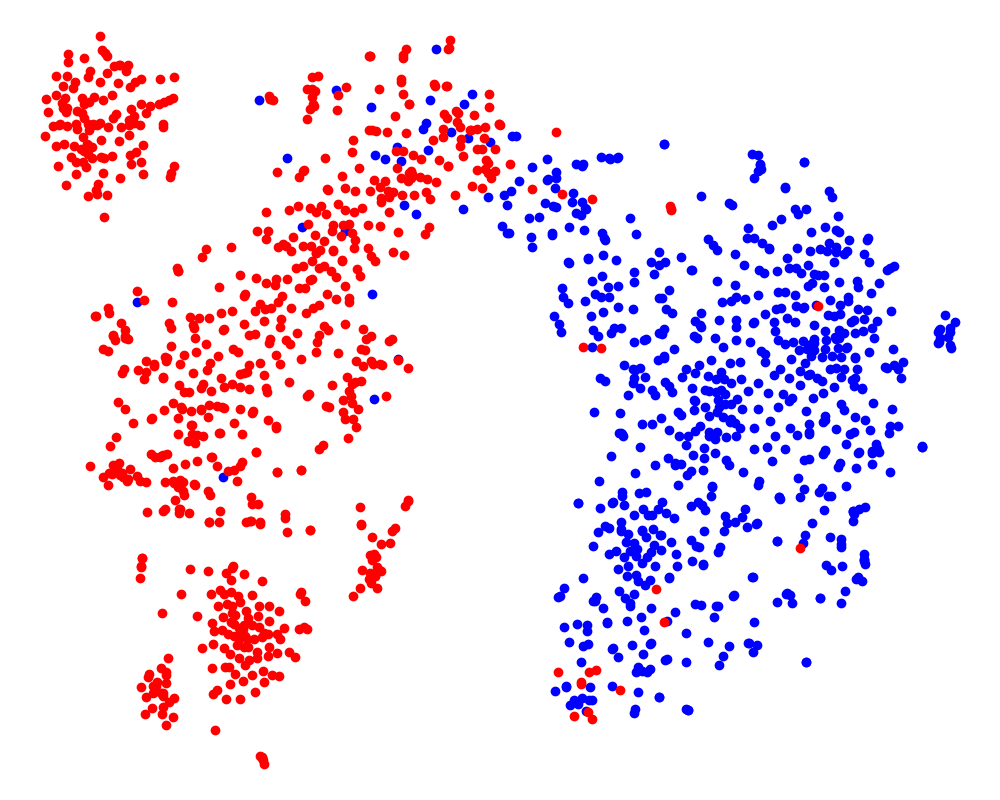}
\caption{w/ MML.}
\label{fig:tsnew/mml}
\end{subfigure}
\centering
\begin{subfigure}[t]{0.192\textwidth}
\centering
\includegraphics[width=3.44cm]{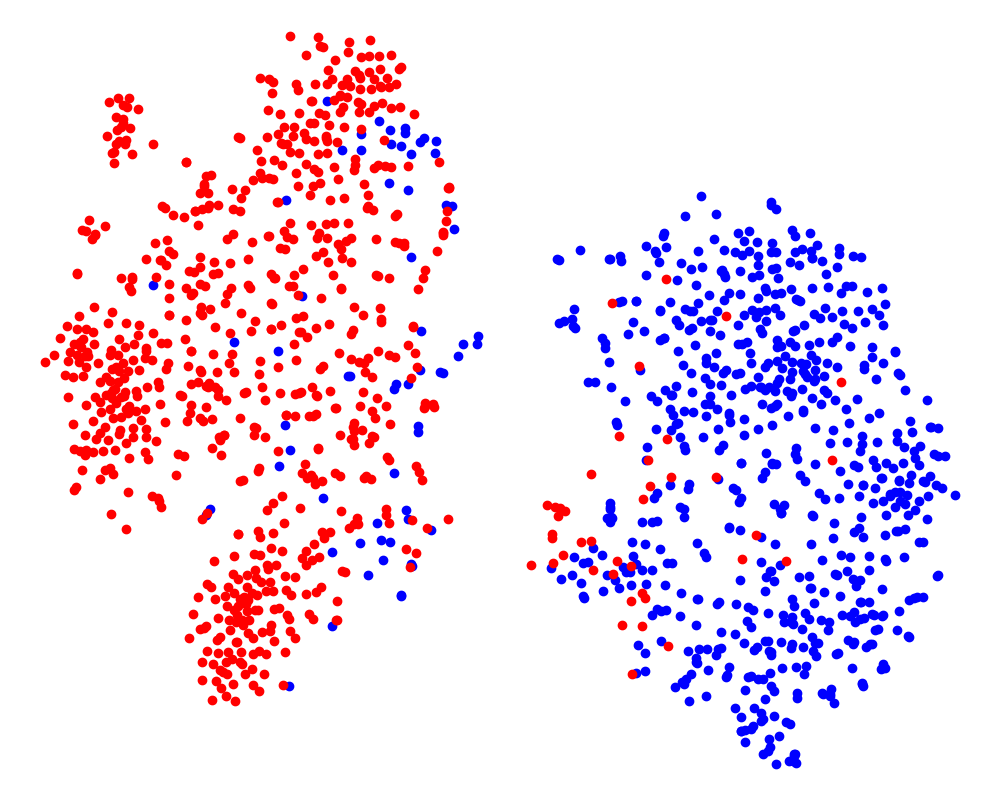}
\caption{w/o Weight.}
\label{fig:tsnew/mmlw/oeq16}
\end{subfigure}
\begin{subfigure}[t]{0.192\textwidth}
\centering
\includegraphics[width=3.44cm]{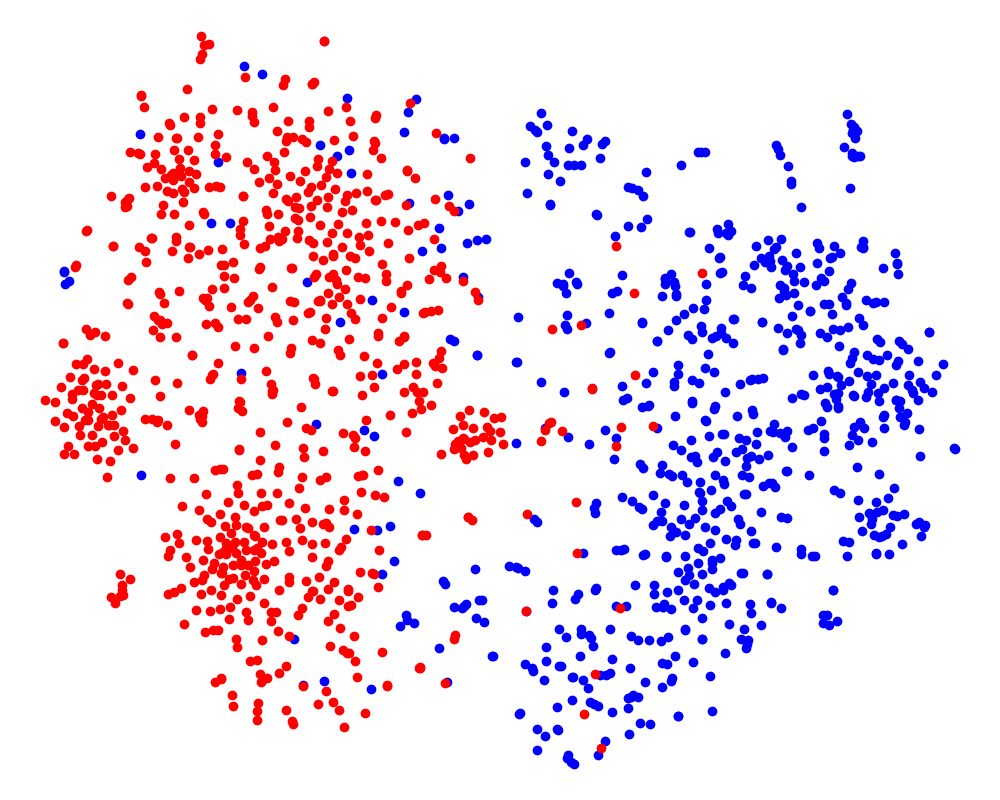}
\caption{w/o MML, w/ MCL.}
\label{fig:tsnew/oMMLw/cl}
\end{subfigure}
\begin{subfigure}[t]{0.192\textwidth}
\centering
\includegraphics[width=3.44cm]{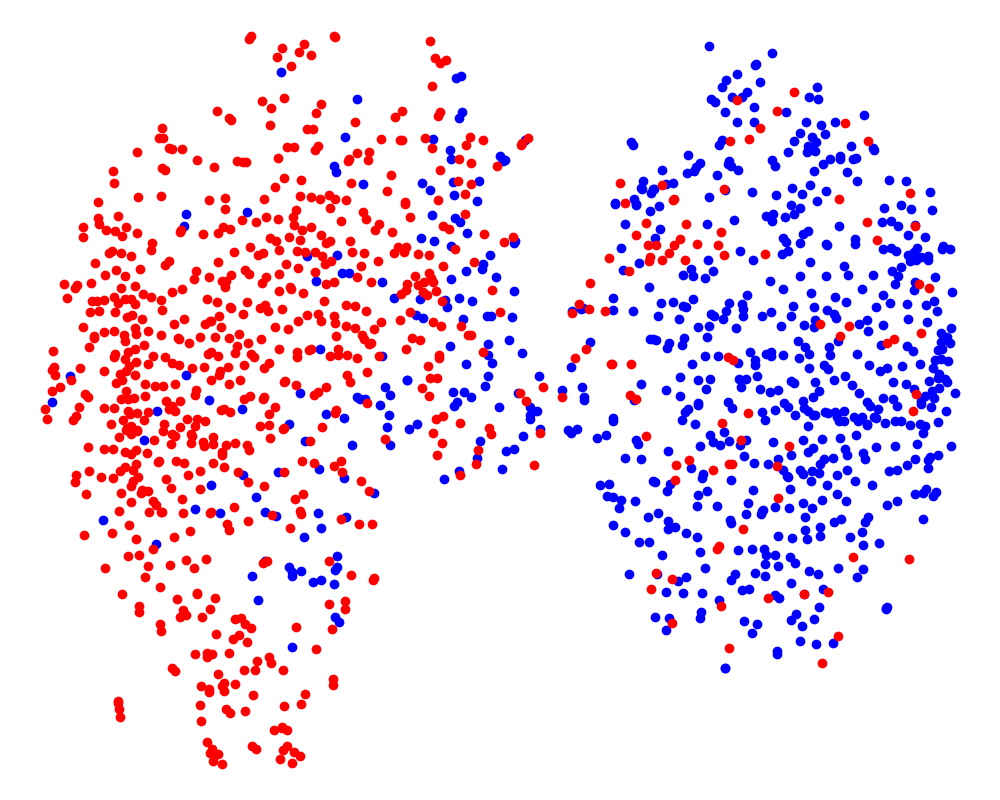}
\caption{w/o MML.}
\label{fig:w/omml}
\end{subfigure}
\caption{Visualization of Ablation Experiments on Modality Missing Learning Module on the Weibo Dataset. Subfigure \ref{fig:curve}: Accuracy on the Testing Set. The Following Four Subfigures: Visualizations of T-SNE Dimensionality Reduction of Multimodal Features Before Classification Corresponding to the Four Curves in Subfigure \ref{fig:curve}.}
\label{fig:mml}
\vspace{-0.20cm} 
\end{figure*}

\subsection{Main Results}

\textbf{Answear to RQ1.} We conducted experiments using Chinese and English datasets in the complete modality scenario, as shown in Table \ref{tab:DROP_xinneng}. MMLNet significantly outperforms existing sota MMR methods and MLLM in three evaluation indicators. On the Chinese weibo dataset, the accuracy scores of the sota MMR method and MLLM are improved by 2.73\% and 12.22\% respectively. On the English pheme dataset, the accuracy scores of the sota MMR method and MLLM are improved by 2.14\% and 30.68\% respectively. These experimental results prove that MMLNet can more accurately identify misinformation in news. For example, the accuracy scores on Weibo and Pheme are 95.22 and 87.78 respectively. This may be attributed to modeling multi-expert modal distribution and jointly discriminating misinformation news. Due to space limitations, the results of the Weibo21 dataset can be viewed in the appendix.

\textbf{Answear to RQ2.} Compared with previous methods of incomplete modality learning \cite{MM24robustSenDistribution}, \cite{MM24robustSenMOE}, \cite{IMOL}, we first explored more fine-grained modality missing scenarios. Inspired by the previous methods, which explored two cases where the image and text modalities were 100\% missing, we designed a total of 14 modality missing scenarios and conducted experiments, covering the previous modality missing settings. Specifically, we set the missing ratio $r$, where the text modality missing ratio is $r_{t}$ and the image modality missing ratio is $r_{v}$. $r = r_{t} + r_{v}$, where $r, r_{t}, r_{v} \in \{0,25,50,75,100\} \subseteq [0,100]$. Here, we visualize the incompleteness of the five modalities when $r=100$, as shown in Figures \ref{fig:weibo}, \ref{fig:weibo21}, and \ref{fig:pheme}. Table \ref{tab:DROP_xinneng} shows in detail the cases where text modality missing dominates ($r_{t}=75$, $r_{v}=25$) and image modality missing dominates ($r_{t}=25$, $r_{v}=75$) when $r=100$. Experiments show that MMLNet performs best in most cases of modality incompleteness and has better robustness. Analyzing the bad cases, in the weibo dataset, when the text modality missing is dominant, although the performance of MIMoE is slightly better than that of MMLNet, when the image modality missing is dominant, the performance of MIMoE deteriorates significantly, with the accuracy dropping by nearly 13\%, while the drop of MMLNet is only 2.6\%. By modeling the distribution of different modalities learned by different experts and learning the consistency of modality missing, MMLNet can cope with various modal missingness with ease. Please refer to the appendix for the complete experiment of modality missing scenario.

\begin{figure*}[t]
  \centering
  \includegraphics[width=\linewidth]{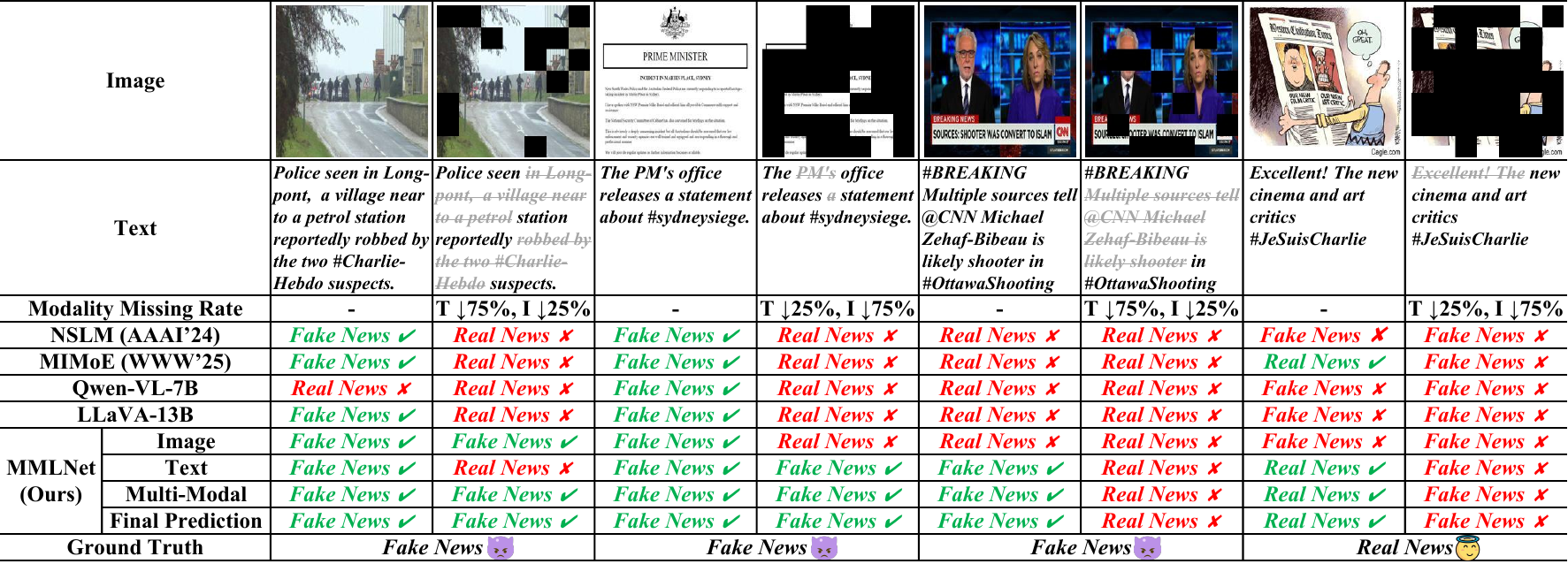}
  \caption{Case study on misinformation recognition. Here, we perform analysis on the Pheme dataset with both complete and incomplete modalities. `T' represents the text modality and `I' represents the image modality. `Text modality missing predominates' refers to T↓75\%, I↓25\%. `Image modality missing predominates' refers to T↓25\%, I↓75\%.}
  \label{fig:case_study}
\end{figure*}

\subsection{Ablation Study}

\textbf{Answear to RQ3.} 
To evaluate the effectiveness of different components, we conduct an ablation study in Table \ref{tab:ab}, which is divided into three parts: 1) modality expert module, 2) modality missing learning module, and 3) incomplete modality adapters. 

1) To validate the effectiveness and robustness of the modality-specfic multi-expert reasoning paradigm in MMLNet for the MFDN task, we conducted ablation studies by sequentially removing the textual expert ${L}_{c}^{h}$, the visual expert ${L}_{c}^{r}$, and the multimodal expert  ${L}_{c}^{f}$. The framework achieved the best performance when all experts collaborated, demonstrating that all modal experts play indispensable roles in discerning the authenticity of news. Notably, removing the ${L}_{c}^{f}$ resulted in a significant performance drop, which highlights the critical importance of modeling the distribution of cross-modal interaction features for the success of MMR.

2) Similarly, we conducted ablation studies by individually removing the textual modality missing learning loss ${L}_{m}^{h}$, visual modality missing learning loss ${L}_{m}^{r}$, and multimodal modality missing learning loss ${L}_{m}^{f}$. Notably, `w/ ${L}_{m}$' indicates that we re-implemented the approach on baseline models NSLM and MIMoE, improved accuracy by an average of 3\% under scenarios where 50\% of textual and visual modalities were missing. Detailed experimental results demonstrate that modality missing learning algorithm enhances robustness. 

3) `w/o $A$' indicates that we removed the adapters that compensate for the missing information of the text and image modalities. `w/ $A$' means that an adapter is added to the baseline model to enhance feature robustness. Interestingly, the adapter not only improves performance in the presence of missing modalities, but also boosts the accuracy score by approximately 1\% even when no modality information is missing, demonstrating that the adapter compensates for missing modality information and enhances the model's robustness.

\subsection{Analysis on Modality Missing Learning.}
\textbf{Answear to RQ4.} As shown in Figure \ref{fig:curve}, our modality missing learning is more suitable for MMR tasks with incomplete modalities. On the Weibo dataset with 50\% text and 50\% image modality missing, we conducted the following ablation experiments on modality missing learning: 
\begin{itemize}
\item \textbf{w/ MML.} MMLNet with modality missing learning.
\item \textbf{w/o Weight.} MMLNet with modality missing learning but without the weight in Eq18.
\item \textbf{w/o MML, w/ MCL.} MMLNet with vanilla multimodal contrastive learning.
\item \textbf{w/o MML.} MMLNet without modality missing learning.
\end{itemize}

We visualized the multimodal features before classification  via T-SNE\footnote{https://github.com/mxl1990/tsne-pytorch} in Figure \ref{fig:mml}.
The \textbf{`w/ MML'} line shows that MMLNet with modality missing learning achieves up to an 4\% higher accuracy than without it, indicating our modality missing learning is beneficial in modality missing scenarios. \textbf{` w/o Weight'} is second-best, showing that positive and negative samples based on news authenticity labels improve the MMLNet model's ability to detect misinformation. \textbf{`w/o MML, w/ MCL'} has similar performance to \textbf{`w/o MML'}, but Figure \ref{fig:tsnew/oMMLw/cl} shows that vanilla multimodal contrastive learning has the worst visualization results, as modality missing undermines vanilla multimodal contrastive learning. This occurs because when image and text modality information is missing, the vanilla multimodal contrastive learning is misled by the missing information and cannot effectively distinguish between positive and negative samples.

\subsection{Case Study}
\textbf{Answear to RQ5.} In Figure \ref{fig:case_study}, we provide a case study on MMR to provide a detailed analysis with some interpretability. Analyzing the left 4 columns, although most baseline models are able to distinguish misinformation when the modality information is complete, they cannot make accurate judgments when the modality information is incomplete. However, MMLNet effectively corrects the impact of missing modality information on model performance by combining distributions from different modality experts. For example, when text modality absence predominates, the lack of text modality information causes erroneous judgments by the text expert, but the image modality expert and multimodal expert distributions compensate for this, improving MMLNet's robustness in MMR.

\textbf{Answear to RQ6.} Although MMLNet has achieved significant improvements in the incomplete modality MMR task, it still has limitations. When modality information is severely missing, it cannot effectively distinguish the authenticity of news. As shown in the sixth column of Figure \ref{fig:case_study}, when the text modallity information is missing to a large extent, the model will mistakenly judge the news as true, mainly because of the loss of key person information. Similarly, when the image modal information is missing to a large extent, effective information cannot be obtained from the image modality. As shown in the eighth column of Figure \ref{fig:case_study}, when the text information on the image is lost, resulting in semantic inconsistency between the image modality and the text modality, the model mistakenly judges the sample as fake news.

\section{Conclusion and Discussion}
In this paper, we focus on the impact of modality missing information on the robustness of multimodal misinformation recognition. These issues are common in practical applications, yet they have been insufficiently explored in previous studies. We propose MMLNet, which models the distribution of different modalities through multi-expert learning. The modality missing adapter compensates for missing modality information, while incomplete modality learning enhances the robustness of feature distributions. Multi-expert collaboration enables accurate identification of misinformation. Extensive experiments and ablation studies under both complete and incomplete modalities demonstrate the effectiveness of our approach. Our future work will explore more missing modality scenarios in MMR tasks, such as video and audio modalities, to further improve the adaptability of our method. Meanwhile, we will study how to design more modality missingness to better fit real-world contexts.

\begin{acknowledgements}
This work was supported by the National Natural Science Foundation of China under Grant No. 62376121 and No. 62501644.
\end{acknowledgements}


\section*{Declarations}

\begin{itemize}
\item \textbf{Conflict of interest}. The authors declare that they have no Conflict of interest.
\item \textbf{Data availability}. The datasets used in this study are publicly available from the following sources: the Weibo dataset \href{https://doi.org/10.1145/3123266.3123454}{\textcolor{blue}{link}}, the Weibo21 dataset \href{https://github.com/\\kennqiang/MDFEND-Weibo21}{\textcolor{blue}{link}}, and the Pheme dataset \href{https://link.springer.com/chapter/10.1007/978-3-319-67217-5_8}{\textcolor{blue}{link}}.
\item \textbf{Code availability}.\\ \textcolor{blue}{https://github.com/zhyhome/MMLNet} 


\end{itemize}


\bibliographystyle{plain}
\bibliography{references}

\clearpage



\onecolumn
\section{Supplementary: Detailed Experimental Results}

\begin{table*}[h]
\small
\centering
\caption{Experimental results of our proposed MMLNet}
\begin{tabular}{cccccccccccc}
\hline
\multirow{2}{*}{Missing Rate (\%)} & \multicolumn{2}{c}{Modality} & \multicolumn{3}{c}{Weibo} & \multicolumn{3}{c}{Weibo21} & \multicolumn{3}{c}{Pheme} \\ \cline{2-12} 
 & Text & Image & ACC & M-F1 & AUC & ACC & M-F1 & AUC & ACC & M-F1 & AUC \\ \hline
\multicolumn{1}{c|}{\multirow{5}{*}{↓100}} & - & \multicolumn{1}{c|}{100} & 91.19 & 91.18 & 96.68 & 90.40 & 90.40 & 95.98 & 83.82 & 79.27 & 90.91 \\
\multicolumn{1}{c|}{} & 25 & \multicolumn{1}{c|}{75} & 90.23 & 90.23 & 96.20 & 90.08 & 90.08 & 95.94 & 82.83 & 79.03 & 86.70 \\
\multicolumn{1}{c|}{} & 50 & \multicolumn{1}{c|}{50} & 91.60 & 91.58 & 96.87 & 89.91 & 89.91 & 96.41 & 82.01 & 76.01 & 74.16 \\
\multicolumn{1}{c|}{} & 75 & \multicolumn{1}{c|}{25} & 92.55 & 92.55 & 96.99 & 90.56 & 90.56 & 96.11 & 80.19 & 74.82 & 84.99 \\
\multicolumn{1}{c|}{} & 100 & \multicolumn{1}{c|}{-} & 92.21 & 92.20 & 97.67 & 89.59 & 89.58 & 95.51 & 80.03 & 73.37 & 85.18 \\ \hline
\multicolumn{1}{c|}{\multirow{4}{*}{↓75}} & - & \multicolumn{1}{c|}{75} & 92.08 & 92.07 & 97.18 & 91.70 & 91.70 & 97.11 & 82.50 & 76.72 & 84.98 \\
\multicolumn{1}{c|}{} & 25 & \multicolumn{1}{c|}{50} & 91.74 & 91.72 & 97.64 & 92.19 & 92.19 & 97.44 & 83.99 & 80.61 & 87.43 \\
\multicolumn{1}{c|}{} & 50 & \multicolumn{1}{c|}{25} & 93.37 & 93.37 & 97.89 & 91.86 & 91.86 & 96.90 & 80.69 & 76.29 & 86.52 \\
\multicolumn{1}{c|}{} & 75 & \multicolumn{1}{c|}{-} & 93.58 & 93.57 & 97.71 & 91.05 & 91.05 & 96.02 & 81.02 & 77.01 & 86.32 \\ \hline
\multicolumn{1}{c|}{\multirow{3}{*}{↓50}} & - & \multicolumn{1}{c|}{50} & 92.96 & 92.96 & 97.92 & 91.70 & 91.70 & 98.03 & 83.49 & 79.48 & 88.42 \\
\multicolumn{1}{c|}{} & 25 & \multicolumn{1}{c|}{25} & 93.92 & 93.92 & 98.64 & 93.33 & 93.33 & 97.69 & 82.17 & 76.39 & 88.18 \\
\multicolumn{1}{c|}{} & 50 & \multicolumn{1}{c|}{-} & 94.40 & 94.39 & 98.50 & 92.19 & 92.19 & 97.50 & 82.50 & 77.76 & 87.60 \\ \hline
\multicolumn{1}{c|}{\multirow{2}{*}{↓25}} & - & \multicolumn{1}{c|}{25} & 94.60 & 94.60 & 98.76 & 93.98 & 93.98 & 98.08 & 83.00 & 77.53 & 90.39 \\
\multicolumn{1}{c|}{} & 25 & \multicolumn{1}{c|}{-} & 94.88 & 94.87 & 98.85 & 93.65 & 93.65 & 97.91 & 84.81 & 79.89 & 88.91 \\ \hline
\multicolumn{1}{c|}{-} & - & \multicolumn{1}{c|}{-} & 95.22 & 95.21 & 98.93 & 94.95 & 94.95 & 98.23 & 87.78 & 84.35 & 90.97 \\ \hline
\end{tabular}
\end{table*}

\begin{table*}[h]
\centering
\small
\caption{Experimental results of NSLM}
\begin{tabular}{cccccccccccc}
\hline
\multirow{2}{*}{Missing Rate (\%)} & \multicolumn{2}{c}{Modality} & \multicolumn{3}{c}{Weibo} & \multicolumn{3}{c}{Weibo21} & \multicolumn{3}{c}{Pheme} \\ \cline{2-12} 
 & Text & Image & ACC & M-F1 & AUC & ACC & M-F1 & AUC & ACC & M-F1 & AUC \\ \hline
\multicolumn{1}{c|}{\multirow{5}{*}{↓100}} & - & \multicolumn{1}{c|}{100} & 88.73 & 88.69 & 89.03 & 89.26 & 89.26 & 89.26 & 78.87 & 75.86 & 77.97 \\
\multicolumn{1}{c|}{} & 25 & \multicolumn{1}{c|}{75} & 86.07 & 85.83 & 85.74 & 84.87 & 84.72 & 84.96 & 82.50 & 80.01 & 82.41 \\
\multicolumn{1}{c|}{} & 50 & \multicolumn{1}{c|}{50} & 82.59 & 82.25 & 82.23 & 84.06 & 83.98 & 84.01 & 80.52 & 75.70 & 75.18 \\
\multicolumn{1}{c|}{} & 75 & \multicolumn{1}{c|}{25} & 71.74 & 70.52 & 71.16 & 77.39 & 77.39 & 77.39 & 74.09 & 60.08 & 59.69 \\
\multicolumn{1}{c|}{} & 100 & \multicolumn{1}{c|}{-} & 48.94 & 48.28 & 49.35 & 53.00 & 45.82 & 53.30 & 71.28 & 41.61 & 50.00 \\ \hline
\multicolumn{1}{c|}{\multirow{4}{*}{↓75}} & - & \multicolumn{1}{c|}{75} & 91.33 & 91.32 & 91.34 & 90.89 & 90.89 & 90.89 & 83.99 & 80.06 & 79.50 \\
\multicolumn{1}{c|}{} & 25 & \multicolumn{1}{c|}{50} & 91.39 & 91.39 & 91.46 & 89.26 & 89.28 & 89.28 & 80.52 & 75.79 & 75.35 \\
\multicolumn{1}{c|}{} & 50 & \multicolumn{1}{c|}{25} & 87.16 & 87.11 & 87.04 & 86.34 & 86.26 & 86.40 & 77.88 & 67.37 & 65.61 \\
\multicolumn{1}{c|}{} & 75 & \multicolumn{1}{c|}{-} & 78.56 & 78.52 & 78.51 & 75.12 & 74.71 & 75.23 & 75.57 & 60.55 & 60.21 \\ \hline
\multicolumn{1}{c|}{\multirow{3}{*}{↓50}} & - & \multicolumn{1}{c|}{50} & 91.67 & 91.67 & 91.73 & 87.64 & 87.58 & 87.59 & 81.18 & 75.89 & 74.79 \\
\multicolumn{1}{c|}{} & 25 & \multicolumn{1}{c|}{25} & 91.19 & 91.18 & 91.15 & 89.59 & 89.57 & 89.63 & 81.84 & 76.83 & 75.77 \\
\multicolumn{1}{c|}{} & 50 & \multicolumn{1}{c|}{-} & 87.03 & 87.01 & 87.01 & 86.66 & 86.63 & 86.71 & 80.03 & 73.37 & 71.75 \\ \hline
\multicolumn{1}{c|}{\multirow{2}{*}{↓25}} & - & \multicolumn{1}{c|}{25} & 92.42 & 92.42 & 92.44 & 90.08 & 90.05 & 90.04 & 82.50 & 77.31 & 75.89 \\
\multicolumn{1}{c|}{} & 25 & \multicolumn{1}{c|}{-} & 91.80 & 91.80 & 91.82 & 90.73 & 90.73 & 90.73 & 81.84 & 78.73 & 80.06 \\ \hline
\multicolumn{1}{c|}{-} & - & \multicolumn{1}{c|}{-} & 92.28 & 92.28 & 92.32 & 90.89 & 90.88 & 90.92 & 84.65 & 81.82 & 82.88 \\ \hline
\end{tabular}
\end{table*}

\begin{table*}[]
\centering
\caption{Experimental results of MIMoE}
\small
\begin{tabular}{cccccccccccc}
\hline
\multirow{2}{*}{Missing Rate (\%)} & \multicolumn{2}{c}{Modality} & \multicolumn{3}{c}{Weibo} & \multicolumn{3}{c}{Weibo21} & \multicolumn{3}{c}{Pheme} \\ \cline{2-12} 
 & Text & Image & ACC & M-F1 & AUC & ACC & M-F1 & AUC & ACC & M-F1 & AUC \\ \hline
\multicolumn{1}{c|}{\multirow{5}{*}{↓100}} & - & \multicolumn{1}{c|}{100} & 90.44 & 90.39 & 90.31 & 88.13 & 88.07 & 88.19 & 80.36 & 72.80 & 87.19 \\
\multicolumn{1}{c|}{} & 25 & \multicolumn{1}{c|}{75} & 90.85 & 90.84 & 91.04 & 89.26 & 89.26 & 89.25 & 77.88 & 67.90 & 83.21 \\
\multicolumn{1}{c|}{} & 50 & \multicolumn{1}{c|}{50} & 84.36 & 84.23 & 84.74 & 86.66 & 86.64 & 86.70 & 73.26 & 58.20 & 79.17 \\
\multicolumn{1}{c|}{} & 75 & \multicolumn{1}{c|}{25} & 80.06 & 80.06 & 80.10 & 82.11 & 82.10 & 82.10 & 74.25 & 61.10 & 75.09 \\
\multicolumn{1}{c|}{} & 100 & \multicolumn{1}{c|}{-} & 74.19 & 73.10 & 73.62 & 79.83 & 79.76 & 79.79 & 72.44 & 55.90 & 72.79 \\ \hline
\multicolumn{1}{c|}{\multirow{4}{*}{↓75}} & - & \multicolumn{1}{c|}{75} & 91.26 & 91.23 & 91.16 & 90.40 & 90.38 & 90.45 & 80.19 & 74.70 & 85.99 \\
\multicolumn{1}{c|}{} & 25 & \multicolumn{1}{c|}{50} & 91.12 & 91.10 & 91.07 & 91.22 & 91.21 & 91.24 & 80.69 & 73.60 & 85.36 \\
\multicolumn{1}{c|}{} & 50 & \multicolumn{1}{c|}{25} & 89.35 & 89.33 & 89.31 & 88.94 & 88.92 & 88.98 & 78.54 & 72.50 & 80.97 \\
\multicolumn{1}{c|}{} & 75 & \multicolumn{1}{c|}{-} & 80.75 & 80.56 & 80.51 & 79.02 & 78.87 & 79.09 & 75.41 & 65.70 & 74.34 \\ \hline
\multicolumn{1}{c|}{\multirow{3}{*}{↓50}} & - & \multicolumn{1}{c|}{50} & 90.92 & 90.88 & 90.80 & 91.38 & 91.38 & 91.39 & 83.16 & 78.40 & 88.21 \\
\multicolumn{1}{c|}{} & 25 & \multicolumn{1}{c|}{25} & 92.69 & 92.69 & 92.73 & 90.89 & 90.88 & 90.92 & 81.35 & 76.30 & 85.80 \\
\multicolumn{1}{c|}{} & 50 & \multicolumn{1}{c|}{-} & 88.80 & 88.80 & 88.85 & 90.56 & 90.56 & 90.55 & 77.22 & 71.40 & 80.75 \\ \hline
\multicolumn{1}{c|}{\multirow{2}{*}{↓25}} & - & \multicolumn{1}{c|}{25} & 93.58 & 93.58 & 93.64 & 92.03 & 92.02 & 92.06 & 82.83 & 79.30 & 88.31 \\
\multicolumn{1}{c|}{} & 25 & \multicolumn{1}{c|}{-} & 92.01 & 92.01 & 92.03 & 92.84 & 92.85 & 92.85 & 82.50 & 79.10 & 87.65 \\ \hline
\multicolumn{1}{c|}{-} & - & \multicolumn{1}{c|}{-} & 92.49 & 92.47 & 92.42 & 91.54 & 91.53 & 91.51 & 85.64 & 81.90 & 89.40 \\ \hline
\end{tabular}
\end{table*}

\begin{table*}[]
\centering
\caption{Experimental results of fine-tuned LLaVA-v1.5-13B}
\small
\begin{tabular}{cccccccccccc}
\hline
\multirow{2}{*}{Missing Rate (\%)} & \multicolumn{2}{c}{Modality} & \multicolumn{3}{c}{Weibo} & \multicolumn{3}{c}{Weibo21} & \multicolumn{3}{c}{Pheme} \\ \cline{2-12} 
 & Text & Image & ACC & M-F1 & AUC & ACC & M-F1 & AUC & ACC & M-F1 & AUC \\ \hline
\multicolumn{1}{c|}{\multirow{5}{*}{↓100}} & - & \multicolumn{1}{c|}{100} & 80.41 & 79.87 & 79.66 & 86.18 & 86.29 & 85.95 & 53.47 & 53.44 & 66.50 \\
\multicolumn{1}{c|}{} & 25 & \multicolumn{1}{c|}{75} & 76.72 & 76.08 & 75.50 & 81.14 & 80.68 & 81.27 & 48.35 & 48.06 & 62.92 \\
\multicolumn{1}{c|}{} & 50 & \multicolumn{1}{c|}{50} & 70.44 & 69.71 & 68.43 & 76.75 & 76.01 & 76.90 & 47.03 & 46.68 & 61.65 \\
\multicolumn{1}{c|}{} & 75 & \multicolumn{1}{c|}{25} & 67.99 & 67.40 & 66.60 & 68.94 & 68.03 & 69.09 & 43.56 & 43.08 & 58.01 \\
\multicolumn{1}{c|}{} & 100 & \multicolumn{1}{c|}{-} & 54.13 & 55.33 & 47.43 & 55.45 & 49.75 & 55.72 & 35.15 & 33.85 & 50.22 \\ \hline
\multicolumn{1}{c|}{\multirow{4}{*}{↓75}} & - & \multicolumn{1}{c|}{75} & 81.71 & 81.25 & 81.17 & 86.67 & 86.47 & 86.67 & 52.48 & 52.43 & 65.64 \\
\multicolumn{1}{c|}{} & 25 & \multicolumn{1}{c|}{50} & 77.34 & 76.76 & 76.35 & 81.30 & 80.83 & 81.43 & 50.17 & 50.03 & 64.02 \\
\multicolumn{1}{c|}{} & 50 & \multicolumn{1}{c|}{25} & 70.99 & 70.26 & 69.04 & 77.72 & 77.09 & 77.86 & 47.36 & 47.09 & 61.53 \\
\multicolumn{1}{c|}{} & 75 & \multicolumn{1}{c|}{-} & 68.12 & 67.65 & 67.17 & 69.92 & 69.17 & 70.05 & 47.19 & 47.15 & 59.19 \\ \hline
\multicolumn{1}{c|}{\multirow{3}{*}{↓50}} & - & \multicolumn{1}{c|}{50} & 82.59 & 82.21 & 82.21 & 86.99 & 86.81 & 87.09 & 54.46 & 54.45 & 66.85 \\
\multicolumn{1}{c|}{} & 25 & \multicolumn{1}{c|}{25} & 77.88 & 76.99 & 77.33 & 81.46 & 80.99 & 81.60 & 51.65 & 51.58 & 65.06 \\
\multicolumn{1}{c|}{} & 50 & \multicolumn{1}{c|}{-} & 71.54 & 70.03 & 70.89 & 77.40 & 76.76 & 77.54 & 48.84 & 48.74 & 62.06 \\ \hline
\multicolumn{1}{c|}{\multirow{2}{*}{↓25}} & - & \multicolumn{1}{c|}{25} & 83.00 & 82.62 & 82.61 & 86.99 & 86.81 & 87.09 & 55.12 & 55.12 & 67.15 \\
\multicolumn{1}{c|}{} & 25 & \multicolumn{1}{c|}{-} & 77.68 & 76.85 & 77.15 & 81.46 & 80.99 & 81.60 & 52.81 & 52.77 & 65.87 \\ \hline
\multicolumn{1}{c|}{-} & - & \multicolumn{1}{c|}{-} & 83.00 & 82.67 & 82.64 & 86.50 & 86.29 & 86.61 & 57.10 & 57.07 & 68.53 \\ \hline
\end{tabular}
\end{table*}

\begin{table*}[]
\centering
\small
\caption{Experimental results of fine-tuned Qwen-VL-7B}
\begin{tabular}{cccccccccccc}
\hline
\multirow{2}{*}{Missing Rate (\%)} & \multicolumn{2}{c}{Modality} & \multicolumn{3}{c}{Weibo} & \multicolumn{3}{c}{Weibo21} & \multicolumn{3}{c}{Pheme} \\ \cline{2-12} 
 & Text & Image & ACC & M-F1 & AUC & ACC & M-F1 & AUC & ACC & M-F1 & AUC \\ \hline
\multicolumn{1}{c|}{\multirow{5}{*}{↓100}} & - & \multicolumn{1}{c|}{100} & 77.88 & 76.60 & 77.21 & 77.07 & 75.93 & 77.26 & 50.17 & 49.99 & 64.36 \\
\multicolumn{1}{c|}{} & 25 & \multicolumn{1}{c|}{75} & 72.90 & 70.60 & 72.08 & 67.32 & 63.87 & 67.57 & 45.87 & 45.37 & 61.01 \\
\multicolumn{1}{c|}{} & 50 & \multicolumn{1}{c|}{50} & 68.67 & 65.31 & 67.74 & 62.76 & 57.52 & 63.05 & 46.04 & 45.62 & 60.78 \\
\multicolumn{1}{c|}{} & 75 & \multicolumn{1}{c|}{25} & 65.46 & 62.32 & 64.60 & 57.40 & 49.97 & 57.72 & 42.08 & 41.45 & 56.80 \\
\multicolumn{1}{c|}{} & 100 & \multicolumn{1}{c|}{-} & 53.65 & 49.24 & 54.65 & 49.76 & 33.51 & 50.16 & 36.80 & 36.51 & 48.46 \\ \hline
\multicolumn{1}{c|}{\multirow{4}{*}{↓75}} & - & \multicolumn{1}{c|}{75} & 79.32 & 78.40 & 78.74 & 77.40 & 76.31 & 77.58 & 49.67 & 49.48 & 63.84 \\
\multicolumn{1}{c|}{} & 25 & \multicolumn{1}{c|}{50} & 74.40 & 72.54 & 73.64 & 68.78 & 65.67 & 69.03 & 48.51 & 48.27 & 62.86 \\
\multicolumn{1}{c|}{} & 50 & \multicolumn{1}{c|}{25} & 68.94 & 65.77 & 68.04 & 64.23 & 59.56 & 64.51 & 47.03 & 46.74 & 61.30 \\
\multicolumn{1}{c|}{} & 75 & \multicolumn{1}{c|}{-} & 66.69 & 64.48 & 65.96 & 56.42 & 48.52 & 56.75 & 45.38 & 45.27 & 57.74 \\ \hline
\multicolumn{1}{c|}{\multirow{3}{*}{↓50}} & - & \multicolumn{1}{c|}{50} & 80.48 & 79.71 & 79.94 & 77.56 & 76.49 & 77.74 & 52.15 & 52.10 & 65.23 \\
\multicolumn{1}{c|}{} & 25 & \multicolumn{1}{c|}{25} & 74.33 & 72.47 & 73.58 & 69.43 & 66.60 & 69.43 & 50.00 & 49.87 & 63.73 \\
\multicolumn{1}{c|}{} & 50 & \multicolumn{1}{c|}{-} & 70.65 & 68.25 & 69.84 & 63.09 & 57.90 & 63.38 & 48.35 & 48.20 & 61.89 \\ \hline
\multicolumn{1}{c|}{\multirow{2}{*}{↓25}} & - & \multicolumn{1}{c|}{25} & 80.89 & 80.20 & 80.37 & 77.40 & 76.31 & 77.58 & 53.14 & 53.12 & 65.93 \\
\multicolumn{1}{c|}{} & 25 & \multicolumn{1}{c|}{-} & 75.09 & 73.52 & 74.38 & 68.62 & 65.53 & 68.87 & 51.32 & 51.26 & 64.48 \\ \hline
\multicolumn{1}{c|}{-} & - & \multicolumn{1}{c|}{-} & 81.71 & 81.12 & 81.22 & 76.91 & 75.74 & 77.09 & 55.12 & 55.12 & 67.15 \\ \hline
\end{tabular}
\end{table*}


\end{document}